\documentclass[a4paper,fleqn]{cas-dc}

\usepackage{ulem}
\usepackage{xcolor}
\usepackage{graphicx,caption,subcaption}
\usepackage{bm}
\usepackage{subcaption}

\newtheorem{assumption}{Assumption}[section]
\usepackage{floatrow}
\newfloatcommand{capbtabbox}{table}[][\FBwidth]
\usepackage[utf8]{inputenc}
\usepackage[english]{babel}
\usepackage{comment}
\usepackage{multirow}
\usepackage{todonotes}

\usepackage[none]{hyphenat}
\usepackage{indentfirst}

\graphicspath{ {./figures/} }
\DeclareGraphicsExtensions{.png}
\usepackage{lipsum}
\usepackage[numbers]{natbib}

\usepackage{listings}
\usepackage{scalerel}
\usepackage[ruled,longend]{algorithm2e}

\newcommand\reallywidehat[1]{\arraycolsep=0pt\relax%
\begin{array}{c}
\stretchto{
  \scaleto{
    \scalerel*[\widthof{\ensuremath{#1}}]{\kern-.5pt\bigwedge\kern-.5pt}
    {\rule[-\textheight/2]{1ex}{\textheight}} 
  }{\textheight} %
}{0.5ex}\\           
#1\\                 
\rule{-1ex}{0ex}
\end{array}
}
\usepackage{amsmath,nccmath}
        \usepackage{setspace}

\def\tsc#1{\csdef{#1}{\textsc{\lowercase{#1}}\xspace}}
\tsc{WGM}
\tsc{QE}
\tsc{EP}
\tsc{PMS}
\tsc{BEC}
\tsc{DE}

\begin{document}

\begin{sloppypar}

\let\WriteBookmarks\relax
\def\floatpagepagefraction{1}
\def\textpagefraction{.001}
\shorttitle{Dynamics and control of clustered tensegrity systems}
\shortauthors{S. Ma, M. Chen, and R. E. Skelton}



\title [mode = title]{Dynamics and control of clustered tensegrity systems}




\author[1]{Shuo Ma}[
orcid=0000-0003-3789-2893]
\fnmark[1]

\address[1]{College of Civil Engineering, Zhejiang University of Technology, Hangzhou, 310014, China}

\author[2]{Muhao Chen}[type=editor,
auid=000,bioid=1,
orcid=0000-0003-1812-6835]
\cormark[1]
\fnmark[2]
\ead{muhaochen@tamu.edu}

\address[2]{Department of Aerospace Engineering, Texas A\&M University, College Station, TX, 77840, USA}

\author[2]{Robert E. Skelton}[orcid=0000-0001-6503-9115]
\fnmark[3]

\cortext[cor1]{Corresponding author. Tel.: +1 9799858285.}

\fntext[fn1]{Assistant Professor, College of Civil Engineering, Zhejiang University of Technology, Hangzhou, 310014, China}
\fntext[fn2]{Postdoc Researcher, Department of Aerospace Engineering, Texas A\&M University, College Station, TX, USA.}
\fntext[fn3]{TEES Eminent Professor, Department of Aerospace Engineering, Texas A\&M University, College Station, TX, USA.}



\maketitle

\begin{abstract}
This paper presents the formulations of nonlinear and linearized statics, dynamics, and control for any clustered tensegrity system (CTS). Based on the Lagrangian method and FEM assumptions, the nonlinear clustered tensegrity dynamics with and without constraints are first derived. It is shown that the traditional tensegrity system (TTS), whose node to node strings are individual ones, yield to be a special case of the CTS. Then, equilibrium equations of the CTS in three standard forms (in terms of nodal coordinate, force density, and force vector) and the compatibility equation are given. Moreover, the linearized dynamics and modal analysis of the CTS with and without constraints are also derived. We also present a nonlinear shape control law for the control of any CTS. The control turns out to be a linear algebra problem in terms of the control variable, which is the force densities in the strings. The statics, dynamics, and control examples are carefully selected to demonstrate the developed principles. The presented approaches can boost the comprehensive studies of the statics, dynamics, and control for any CTS, as well as promoting the integration of structure and control design.
\end{abstract}

\section{Introduction}

Biological systems perhaps provide the greatest evidence that tensegrity systems are the most efficient structures. For example, living cells use microtubules and micro-filaments to control their surfaces \cite{wang2001mechanical}. The electron micrographs show that the DNA bundles are consistent with the tensegrity prism unit \cite{liedl2010self}. The elbow of Humans and animals are also tensegrity structures. After years of study, tensegrity has shown its many advantages in lightweight structure designs \cite{ma2020design,wang2018wave,wang2021form}, high stiffness-to-mass ratio \cite{feng2018dynamic,fraddosio2019minimal,wang2021minimal}, and structure deployability \cite{sychterz2018using,yang2019deployment}. For example, Ma \textit{et al.} designed a mass efficient tensegrity cantilever structure with and without wall-length constraints \cite{ma2020design}. 
Branam \textit{et al.} presented nonlinear static analysis of cable structures and for the form-finding of tensegrity structures
\cite{branam2019unified}. Wang \textit{et al.} presented a mass design approach to active tensegrity structures \cite{wang2021minimal}. Shintake \textit{et al.} designed a 400 mm long tensegrity fish-like robots driven with tunable stiffness \cite{shintake2020bio}. Zhou \textit{et al.} built and investigated a prestressable tensegrity morphing airfoil with six pneumatic actuators \cite{zhou2021distributed}. Wang \textit{et al.} constructed a hybrid tensegrity robot composed of hard and soft materials, mimicking the musculoskeletal system of animals \cite{wang2019light}. Veuve \textit{et al.} presented measurements and control methodologies for a deployable tensegrity structure based on an efficient learning strategy and a damage-compensation algorithm \cite{veuve2017adaptive}. Furthermore, to date, a few pieces of research on meta-material-based tensegrity structures have been conducted. For example, Lee \textit{et al.} reported an approach to fabricate tensegrity structures by smart materials using 3D printing and sacrificial molding \cite{lee20203d}. Rimoli and Pal studied the efficient modeling of the constitutive behavior of tensegrity meta-materials under a wide range of loading conditions and prestressed configurations \cite{rimoli2017mechanical}. 

Tensegrity is a stable network of compressive members (bars/strut) and tensile members (strings/cables) \cite{skelton2009tensegrity}. The clustered tensegrity is a tensegrity structure with clustered strings. A clustered string is a group of individual cables that are combined into one continuous string that runs over frictionless pulleys or through frictionless loops at the nodes \cite{moored2009investigation}. Since the clustered actuation strategy can reduce the number of actuators, sensors, and related devices, it has been studied by various researchers for deployable pantographic structures  \cite{kwan1994matrix,kwan1993active} and deployable tensegrities \cite{smaili2005folding,veuve2015deployment}.

A few pieces of research have been conducted on CTS. For example,  Moored and Bart-Smith investigated the actuation strategies and prestress mechanism and stability analysis of the clustered tensegrity structures \cite{moored2009investigation}. Kan \textit{et al.} presented a sliding cable element for multibody dynamics with an application to the deployment of clustered tensegrity by using the configuration of the attached rigid bodies as the generalized coordinates \cite{Kan2017A}. Kan \textit{et al.} derived the dynamic analysis of clustered tensegrity structures via the framework of the positional formulation FEM \cite{kan2018nonlinear}. Ali \textit{et al.} presented the static analysis and form-finding problems of the clustered tensegrity structures using a modified dynamic relaxation algorithm \cite{ali2011analysis}. Zhang \textit{et al.} presented a FEM formulation for a geometrically nonlinear elasto-plastic analyses approach based on the co-rotational approach \cite{zhang2015geometrically}. Shuo \textit{et al.} designed and analyzed a clustered deployable tensegrity cable dome \cite{ma2021design}. However, few of these provided a general compact form of nonlinear dynamics (allowing structure members to have both elastic or plastic deformations), linearized dynamics as we as a control law for any CTS. In this paper, we presented an explicit form of nonlinear and linearized dynamics and a closed-loop control law for the insight knowledge of the clustered tensegrity dynamics and future convenience for the field of structural control.

This paper is organized as follows: Section \ref{section2} describes bar and string assumptions, nodal coordinates and connectivity matrix notations, and geometric and physical properties of the CTS in compact vector forms. Section \ref{section3} formulates the shape function of a structural element, kinetic energy, strain, and gravitational potential energy of the whole structure. Then, clustered tensegrity dynamics with and without boundary constraints are derived by the Lagrangian method. By neglecting the time derivative terms in the dynamics equation, Section \ref{section4} gives the equilibrium equations of the CTS in three standard equivalent forms (in terms of nodal vector, force density, and force vectors) and the compatibility equation. Section \ref{section5} derives the linearized clustered tensegrity dynamics and modal analysis equations with and without boundary constraints. Section \ref{section6} presents a nonlinear control law for the control of any CTS. Section \ref{section7} demonstrates three examples to verify the accuracy and efficiency of the proposed dynamics equations and the control law. Section \ref{section8} summarises the conclusions.

\section{Notations of the clustered tensegrity system}
\label{section2}

\subsection{Assumptions of structural members}

The clustered tensegrity dynamics is composed of bars, strings, and joints, which follow the list assumptions:
\begin{assumption}
The structural members (bars, strings, and joints) in the clustered tensegrity have these properties: \\
\textbf{Bars}: 
1). Axially loaded. 2). Not rigid, allowed to have elastic or plastic deformation. 3). Negligible inertia about their longitudinal axes. 4). Homogeneous along their length and mass is evenly distributed along the bar.\\
\textbf{Strings}: 1). Axially loaded. 2). Allowed to have elastic or plastic deformation. 3). Negligible inertia about their longitudinal axes. 4). Homogeneous along their length and mass is evenly distributed along the string. 5). A string can never push along its length. If so, the tension in the string should be substituted for zero. 6). The clustered strings are connected by pulleys. 7). All the segments in one clustered string have the same tension.\\
\textbf{Joints}: 1). The bar-bar, bar-string, and non-clustered string-string joints are negligibly small and frictionless pin joints. 2). The clustered string-string joints are negligibly small and frictionless pulleys. 
\end{assumption}

\subsection{Generalized coordinates and configuration}
\label{Generalized}

The nodal coordinates of bars and strings determine the configuration of a tensegrity structure. Let us define a nodal coordinate vector $\bm{n} \in \mathbb{R}^{3 n_n}$ to describe the structure:
\begin{align}
    \bm{n} = \begin{bmatrix} \bm{n}_1^T & \bm{n}_2^T & \cdots & \bm{n}_{n_n}^T\end{bmatrix}^T,
\end{align}
where $\bm{n}_i = \begin{bmatrix}  x_i & y_i & z_i\end{bmatrix} ^T  \in \mathbb{R}^3$ is the $x$-, $y$-, and $z$-coordinate vector of the \textit{i}th node, $i = 1, 2, \cdots, n_n$. $n_n$ is the number of nodes of the structure. The nodal coordinate of the structure can also be written into a nodal matrix $\bm{N}\in \mathbb{R}^{3\times n_n}$:
\begin{align}
   \bm{N} = \begin{bmatrix}
    \bm{n}_1 & \bm{n}_2 & \cdots & \bm{n}_{n_n}
    \end{bmatrix}.
\end{align}

For many practical problems, we have to restrict the motion of the structure in certain directions by fixing or giving the position, velocity, or acceleration values of some nodes in the structure. The constraints will reduce the degree of freedom of the dynamics in a smaller space. Thus, to describe the reduced-order dynamics, we define two vectors $\bm{a}\in \mathbb{R}^{n_{a}} $ and $\bm{b}\in \mathbb{R}^{n_{b}}$ to denote the indices of the free and constrained nodal coordinates:
\begin{align}\label{vec_a}
    & \bm{a}=\begin{bmatrix}a_{1} & a_{2}& \cdots & a_{n_{a}}\end{bmatrix}^{T}, \\  \label{vec_b}
    & \bm{b}=\begin{bmatrix}b_{1}& b_{2}& \cdots & b_{n_{b}}\end{bmatrix}^{T},
\end{align}
where $n_a$ and $n_b$ are the number of free and constrained nodal coordinates ($n_a+n_b=3n_n$), $a_i$ and $b_i$ are the indices of free and constrained entries in the nodal coordinate vector $\bm{n}$. Let index matrices $\bm{E}_{a} \in \mathbb{R}^{3 n_{n} \times n_{a}}$ and $\bm{E}_{b} \in \mathbb{R}^{3 n_{n} \times n_{b}}$ be related to the Eqs. (\ref{vec_a}) and (\ref{vec_b}):
\begin{align}
\bm{E}_{a}\left(:, i\right)=\textbf{I}_{3 n_n}\left(:, a_{i}\right), ~\bm{E}_{b}\left(:, i\right)=\textbf{I}_{3 n_n}\left(:, b_{i}\right),
\end{align}
where $\textbf{I}_{3 n_n}$ is the identity matrix with $3 n_n$ order. Then, the free and constrained nodal coordinate vectors $\bm{n}_a$ and $\bm{n}_b$ with respect to $\bm{n}$ can be obtained:
\begin{align}
\bm{n}_{a}=\bm{E}_{a}^{T} \bm{n},~ \bm{n}_{b}=\bm{E}_{b}^{T} \bm{n}.
\label{na}
\end{align}
Since $\begin{bmatrix}\bm{E}_{a} &  \bm{E}_{b}\end{bmatrix}$ is an orthogonal matrix, we have the following equation:
\begin{align}\bm{n}=
\begin{bmatrix}
\bm{E}_{a}^{T} \\
\bm{E}_{b}^{T}
\end{bmatrix}^{-1}
\begin{bmatrix}\bm{n}_{a} \\
\bm{n}_{b}
\end{bmatrix}=
\begin{bmatrix}
\bm{E}_{a} & \bm{E}_{b}\end{bmatrix}
\begin{bmatrix}
\bm{n}_{a} \\
\bm{n}_{b}
\end{bmatrix}.
\label{n=[Ea Eb][na;nb]}
\end{align}




\subsection{Connectivity matrix}

A connectivity matrix denotes the network pattern of the bars and strings in a tensegrity system. To provide the freedom of clustering any cluster-able strings in a TTS, whose strings are individual ones, we first give the connectivity matrix of the TTS. Then, a clustering matrix is defined to label the connectivity information of clustering strings, which will be discussed in Section \ref{cluster_matrix_sec}.

Let $\bm{C} \in \mathbb{R}^{n_e \times n_n}$ be the connectivity matrix of a TTS, where $n_e$ is the sum of bars and strings. The $m$th ($m= 1, 2, \cdots, n_e$) row of $\bm{C}$,  denoted as $\bm{C}_m = [\bm{C}]_{(m,:)} \in \mathbb{R}^{1\times n_n}$, represents the $m$th element in the TTS, which is described by a vector starting from node $j$ ($j = 1, 2, \cdots, n_n$) to node $k$ ($k = 1, 2, \cdots, n_n$). The $i$th ($i= 1, 2, \cdots, n_n$) entry of $\bm{C}_{m}$ satisfies:
\begin{align}
[\bm{C}]_{mi}=\left\{
\begin{aligned}
-1 &,~ i=j\\
1 &,~ i=k\\
0 &,~ i=\text{else} 
\end{aligned}
\right..
\end{align}
The overall structure connectivity matrix of the TTS $\bm{C}\in  \mathbb{R}^{n_e \times n_n}$ is:
\begin{align}
\bm{C} = \begin{bmatrix} \bm{C}_1^T & \bm{C}_2^T & \cdots & \bm{C}_{n_e}^T\end{bmatrix} ^T .
\end{align}

We define a nodal coordinate vector of the \textit{m}th ($m= 1, 2, \cdots, n_e$) element $\bm{n}_m^e  \in \mathbb{R}^6$ in the TTS as:
\begin{align}\label{n_me}
\bm{n}_m^e=\begin{bmatrix}
\bm{n}_j \\
\bm{n}_k \\
\end{bmatrix}=\begin{bmatrix} x_j & y_j& z_j&x_k&y_k&z_k\end{bmatrix}^T,
\end{align}
which can be written in terms of the structure nodal coordinate vector $\bm{n}$:
\begin{align}
    \bm{n}_m^e=\Bar{\bm{C}}_m\otimes \textbf{I}_3\bm{n},
    \label{n_i^e}
\end{align}
where $\textbf{I}_{3} \in \mathbb{R}^{3\times 3}$ is a identity matrix, $\Bar{\bm{C}}_m$ is a transformation matrix, whose \textit{p}th column satisfies:
\begin{align}
[\bar{C}_{m}]_{(:,p)}=\left\{\begin{array}{ll}
{\begin{bmatrix}1&0\end{bmatrix}^T}, & p=j \\
{\begin{bmatrix}0&1\end{bmatrix}^T}, & p=k \\
{\begin{bmatrix}0&0\end{bmatrix}^T}, & p=\text{else}
\end{array}\right..
\end{align}

\subsection{Clustering matrix}
\label{cluster_matrix_sec}


For easy deployability of the tensegrity structure (i.e., use fewer actuators to pull the strings), some adjacent strings in the TTS can be connected as a single string (also called a clustered string) running over the string-string joints by pulleys \cite{moored2009investigation}. Recall that in TTS, the number of segments is $n_e$. For the CTS, we denote the number of elements as $n_{ec}$, including bars, non-clustered, and clustered strings. If we use clustered strings in a TTS, we have $n_{ec} < n_e$. And if $n_{ec}=n_e$, the clustered tensegrity structure yields into a traditional one. In other words, the TTS is a special case of the clustered one.

Now, we introduce a clustering matrix $\bm{\mathcal{S}}\in \mathbb{R}^{n_{ec} \times n_e}$ to label the connectivity information of the clustered strings:
\begin{align}
\bm{[\mathcal{S}]}_{ij}=\left\{
\begin{aligned}
1 &,~\text{if the \textit{i}th clustered element is }
\\ & ~\text{ composed
of the \textit{j}th classic element.}\\
0 &,~\text{otherwise.}
\end{aligned}
\right.
\end{align}
One can observe that, if $\mathcal{S}=\bm{I}$, the clustered tensegrity structure is equivalent to a traditional one.

\subsection{Geometric properties of the structural elements}

\begin{figure}
    \centering
    \includegraphics[scale=0.8]{ 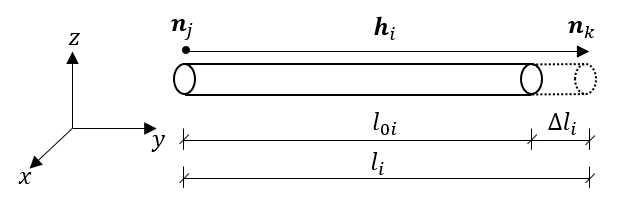}
    \caption{The $i$th structure element $\bm{h}_i$, determined by node $\bm{n}_j$ and node $\bm{n}_k$ in the Cartesian coordinates, has a length of $l_i  = ||\bm{h}_i||= l_{0i} + \Delta l_i$, where $l_{0i}$ is the rest length and $\Delta l_i$ is the displacement.}
    \label{geometry}
\end{figure}

Similarly, we give the notations of the TTS first. Then, by using the clustering matrix  $\mathcal{S}$, the geometric properties of the clustered tensegrity structure are formulated. 

Generally, a structure element can be determined by the two nodal coordinates at its two ends, for example, the $m$th structure element $\bm{h}_m$ can be written as:
\begin{align}
    \bm{h}_m=\bm{n}_k-\bm{n}_j=\bm{C}_m\otimes\textbf{I}_3\bm{n},
\end{align}
and its length is:
\begin{align}
    l_m=\lVert \bm{h}_m\rVert=(\bm{n}^T(\bm{C}_m^T\bm{C}_m)\otimes\textbf{I}_3\bm{n})^\frac{1}{2}.
    \label{li}
\end{align}

The structure element matrix $\bm{H}\in \mathbb{R}^{3\times n_e}$, structure element length vector $\bm{l} \in \mathbb{R}^{n_e}$, and rest length vector $\bm{l}_0 \in \mathbb{R}^{n_e}$ of the TTS can be written as:
\begin{align}
    \bm{H} & =\begin{bmatrix}\bm{h}_1 & \bm{h}_2&\cdots&\bm{h}_{n_e}\end{bmatrix}=\bm{NC}^T,\\
    \bm{l} & =\begin{bmatrix}l_1&l_2&\cdots&l_{n_e}\end{bmatrix}^T,\\
    \bm{l}_0 & =\begin{bmatrix} l_{01}&l_{02}&\cdots&l_{0{n_e}}\end{bmatrix}^T, 
\end{align}
where rest length is the length of an structure element with no tension or compression. 

For a clustered tensegrity, the element length vector $\bm{l}_c\in \mathbb{R}^{n_{ec}}$ is \cite{ ali2011analysis,moored2009investigation}:
\begin{equation}
    \bm{l}_c=\bm{\mathcal{S}}\bm{l}.
    \label{l_c=Sl}
\end{equation}
Similarly, the elements' rest length vector $\bm{l}_{0c} \in \mathbb{R}^{n_{ec}}$ of the CTS is:
\begin{equation}
    \bm{l}_{0c} =\bm{\mathcal{S}}\bm{l}_0.
\end{equation}


\subsection{Physical properties of the structural elements}
\begin{figure}
    \centering
    \includegraphics[width=2.5in]{ 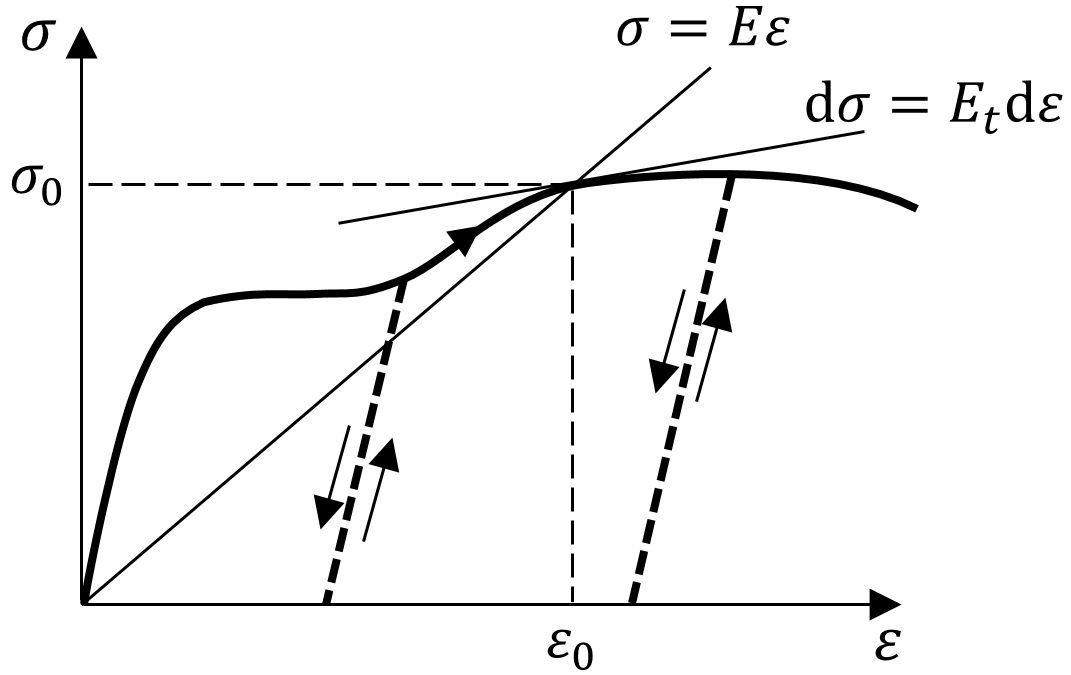}
    \caption{{{}The solid line is a typical stress-strain curve of structure materials, where $E$ and $E_t$ are secant and tangent modulus of the material. The dotted lines are the stress-strain levels for unloading cases at certain points of the stress-strain curve.}}
    \label{stress strain curve}
\end{figure}


For many engineering applications, one may use the elastic and plastic deformation properties of the materials to balance the cost and structural strength. Take the structure performance in earthquake-resistant capacities as an example; normally, the design criteria is the structure will not yield, is repairable, and will not fall under small, medium, and large vibrations. Thus, in this dynamics formulation, we allow the structural members to have elastic or plastic deformation, and a general stress-strain curve of an element is shown in Fig.\ref{stress strain curve}. 

The stress in the structural element satisfies:
\begin{equation}
    \sigma=E\epsilon,
    \label{sigma}
\end{equation}
where $E$ is the secant modulus and $\epsilon$ is the strain, and this equation can represent the stress of any material, i.e., linear elastic, multi-linear elastic, plastic. The derivative of Eq. (\ref{sigma}) is: 
\begin{equation}
       \mathrm{d} \sigma=E_t\mathrm{d} \epsilon,
    \label{d sigma}
\end{equation}
where $E_t$ is the tangent modulus, for linear elastic material, the secant modulus is identical to its tangent modulus. And for multi-linear elastic and plastic materials, the relationship between the secant modulus and tangent modulus is shown in Fig.\ref{stress strain curve}. 

Let the cross section area, secant modulus, tangent modulus of the \textit{i}th structure element be $A_i$, $E_i$, and $E_{ti}$, and the material density be $\rho$, the element mass $m_i$ satisfies $m_i =\rho A_i l_{0i}$. Denote the cross-section area, secant modulus, tangent modulus, and mass vectors of the CTS as $\bm{A}_c$, $\bm{E}_c$, $\bm{E}_{ct}, \bm{m}_c \in \mathbb{R}^{{n}_{ec}}$:
\begin{align}
       \bm{A}_c & =\begin{bmatrix}A_{c1}& A_{c2}&\cdots&A_{cn_{ec}}\end{bmatrix}^T,\\
   \bm{E}_c & =\begin{bmatrix}E_{c1}& E_{c2}&\cdots&E_{cn_{ec}}\end{bmatrix}^T,\\
   \bm{E}_{tc} & =\begin{bmatrix}E_{tc1}& E_{tc2}&\cdots&E_{tn_{ec}}\end{bmatrix}^T,\\
       \bm{m}_c & =\begin{bmatrix}m_{c1}& m_{c2}&\cdots&m_{cn_{ec}}\end{bmatrix}^T=\rho\hat{\bm{A}}_c\bm{l}_{0c},
\end{align}
where $\hat{\bm{v}}$ transforms a vector $\bm{v}$ into a diagonal matrix, whose diagonal entries are the elements of vector $\bm{v}$ and elsewhere are zeros. The cross sectional area, secant modulus, tangent modulus and mass vector of the corresponding TTS is $\bm{A}$, $\bm{E}$, $\bm{E}_t,\bm{m}\in \mathbb{R}^{n_e}$, we also have:
\begin{align}
      \bm{A}& =\begin{bmatrix}A_{1}&A_{2}&\cdots&A_{n_e} \end{bmatrix}^T=\bm{\mathcal{S}}^T \bm{A}_c,\\
   \bm{E} & =\begin{bmatrix}E_{1}&E_{2}&\cdots&E_{n_e} \end{bmatrix}^T=\bm{\mathcal{S}}^T \bm{E}_c,\\
   \bm{E}_t &=\begin{bmatrix}E_{t1}&E_{t2}&\cdots&E_{tn_e}
   \end{bmatrix}^T=\bm{\mathcal{S}}^T \bm{E}_{tc},\\
    \bm{m} & =\begin{bmatrix}m_{1}&m_{2}&\cdots& m_{n_e}\end{bmatrix}^T=\rho\hat{\bm{A}}\bm{l}_0.
    \label{mass vector}
\end{align}


The internal force of the \textit{i}th element of the CTS is $t_{ci} =A_{ci}\sigma_{ci}= E_{ci}A_{ci}(l_{ci}-l_{0ci})/l_{0ci}$. And the internal force vector of CTS and TTS can be written as:
\begin{align}
    \bm{t}_c&=   \hat{\bm{E}}_c\hat{\bm{A}}_c\hat{\bm{l}}_{0c}^{-1}(\bm{l}_c-\bm{l}_{0c})
    \label{t_c force vector},    \\ 
    \bm{t}&=\hat{\bm{E}}\hat{\bm{A}}\hat{\bm{l}}_0^{-1}(\bm{l}-\bm{l}_0)=\bm{\mathcal{S}}^T\bm{t}_c\label{t force vector}.
\end{align}

The force density (force per unit length) of the \textit{i}th element in the CTS is $x_{ci}=t_{ci}/l_{ci}$. Then, the force density vectors of the CTS and TTS are:
\begin{align}
\bm{x}_c&=\hat{\bm{l}}_c^{-1} \bm{t}_c=\hat{\bm{E}}_c \hat{\bm{A}}_c(\bm{l}_{0c}^{-1}-\bm{l}_c^{-1}) 
\label{x_bar force density},\\
\bm{x}&=\hat{\bm{E}}\hat{\bm{A}}(\bm{l}_0^{-1}-\bm{l}^{-1})=\hat{\bm{l}}^{-1}\mathcal{S}^T\bm{t}_c=\hat{\bm{l}}^{-1}\bm{\mathcal{S}}^T\hat{\bm{l}}_c\bm{x}_c,
\label{x force density}
\end{align}
where $\bm{\bm{v}}^{-1}$ represents a vector whose entry is the reciprocal of its corresponding entry in $\bm{v}$. 




\section{Nonlinear tensegrity dynamics formulation}\label{section3}

In this section, we formulated the kinetic and strain potential and gravitational potential energy functions of the CTS and derived the dynamics by the Lagrangian method.

\subsection{The Lagrangian method}

The general form of the Lagrangian equation is:
\begin{align}
    \frac{\mathrm{d}}{\mathrm{d}t}\frac{\partial L}{\partial \dot{\bm{q}}}-\frac{\partial L}{\partial \bm{q}}=\bm{f}_{np},
    \label{Lagarangian's equation}
\end{align}
where $L=T-V$ is the Lagrangian function, $T$ and $V$ are the kinetic energy and potential energy of the system, $\bm{f}_{np}$ is the non-potential force vector on the nodes of the tensegrity structures, and $\bm{q}$ is the generalized coordinate of the system, which is the free nodal coordinate vector $\bm{n}_a$ in our derivation. The potential energy of the whole structure is the sum of strain energy $V_e$ and gravitational potential energy $V_g$:
\begin{align}
    V=V_e+V_g.
\end{align}
Then, the Lagrangian's function can be written as:
\begin{align}
    L=T-(V_e+V_g).
    \label{L}
\end{align}

\label{Seciton 3}
\subsection{Energy equation formulation}


The kinetic energy, strain and gravitational potential energy of the systems are related to the energy of the particles in the structure elements, which are independent of the clustering strategies. Thus, the formulation of the energy functions of CTS and TTS are the same. In the following subsections, we first use the TTS notions to do the derivation, and then by using the clustering matrix, we convert the TTS equations to the CTS ones. 


\subsubsection{Shape function of the structure element}


\begin{figure}
    \centering
    \includegraphics[scale=0.35]{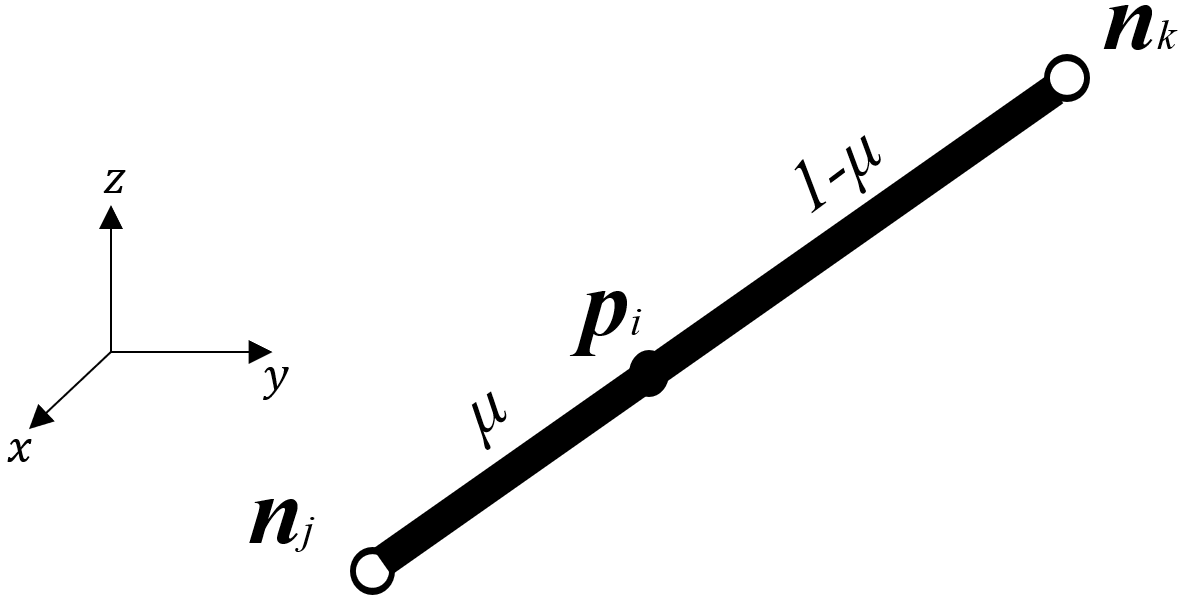}
    \caption{Shape function of a structure element.}
    \label{shape function}
\end{figure}








From the definition of tensegrity structures, we know the structure members are all axially loaded, so the displacement of the material particles is along the bar/string vectors. We assume the displacement of material particles in the structural member is in a uniform manner \cite{bathe2007finite}. So, by knowing the coordinates of one structure member at its two ends, we can interpolate the coordinates of every particle in the structure member. 

Mathematically, we introduce a scalar $\mu$ to help expressing the coordinates of point $\bm{p}_i(\mu)$ on the $i$th member between node $\bm{n}_j$ and node $\bm{n}_k$ in the \textit{i}th structure element, as shown in Fig.\ref{shape function}. Thus, the location of a point $\bm{p}_i(\mu)$ on the structural member can be written as a linear function in terms of $\mu$:
\begin{align}    
    \bm{p}_i(\mu)&=\begin{bmatrix} 1-\mu & \mu\end{bmatrix}\otimes\textbf{I}_3\begin{bmatrix}\bm{n}_j \\\bm{n}_k\end{bmatrix}=\bm{N}^e \bm{n}_i^e, \label{r=N^en_i^e} \\
        \bm{N}^e&=\begin{bmatrix}1-\mu&\mu\end{bmatrix}\otimes \textbf{I}_3,
\end{align}
where $\bm{N}^e\in \mathbb{R}^{3\times 6}$ is usually called the shape function of a structure element.
\subsubsection{Kinetic energy}


As we assume the pulleys are negligibly small, the moment of inertia of the pulleys can also be neglected. That is, the kinetic energy of the structure $T$ is only the sum of the kinetic energy of all the moving particles in the structure members, which is a function of the particle velocity $\dot{\bm{p}}_i(\mu)$:
\begin{align}
    \label{T1}
    T=\sum\limits_{i=1}^{n_e}\frac 12\int\limits_{0}^1m_i\lVert\dot{\bm{p}}_i(\mu)\rVert^2d\mu. 
\end{align}
Substitute Eq. (\ref{n_i^e}) and Eq. (\ref{r=N^en_i^e}) into Eq. (\ref{T1}), we have:
\begin{align}
    T &= \sum\limits_{i=1}^{n_e}\frac 12\int\limits_{0}^1m_i(\bm{N}^e\bar{\bm{C}}_i\otimes \textbf{I}_3\dot{\bm{n}})^2d\mu\\
    &=\sum\limits_{i=1}^{n_e}\frac {m_i}{12}\dot{\bm{n}}^T(\bar{\bm{C}}_i^T\begin{bmatrix}2&1\\1&2\end{bmatrix}  \bar{\bm{C}}_i)\otimes \textbf{I}_3\dot{\bm{n}} \\
    &=\sum\limits_{i=1}^{n_e}\frac {m_i}{12}\dot{\bm{n}}^T(\bar{\bm{C}}_i^T(\begin{bmatrix}1\\1\end{bmatrix}[1~1]+ \lfloor\begin{bmatrix}1\\1\end{bmatrix}[1~1]\rfloor )\bar{\bm{C}}_i)\otimes \textbf{I}_3\dot{\bm{n}} \\
    &=\sum\limits_{i=1}^{n_e}\frac {1}{12}\dot{\bm{n}}^T(|\bm{C}|_i^Tm_i|\bm{C}|_i+ \lfloor |\bm{C}|_i^Tm_i |\bm{C}|_i\rfloor)\otimes \textbf{I}_3 \dot{\bm{n}} \\   
   &=\frac {1} {12}\dot{\bm{n}}^T(|\bm{C}|^T\hat{m}|\bm{C}|+ \lfloor|\bm{C}|^T\hat{m}|\bm{C}|\rfloor)\otimes \textbf{I}_3\dot{\bm{n}}\\
   &=\frac {1} {2}\dot{\bm{n}}^T \bm{M}\dot{\bm{n}},
\end{align}
where $|\bm{V}|$ is an operator that gets the absolute value of each element of a given matrix $\bm{V}$, and the operator $\lfloor \bm{V} \rfloor$ sets all the off-diagonal elements of a square matrix to zero. $\bm{M}\in \mathbb{R}^{3n_n\times 3n_n}$ is called the mass matrix of the structure:
\begin{align}
   \bm{M} =\frac {1} {6}(|\bm{C}|^T\hat{\bm{m}}|\bm{C}|+ \lfloor|\bm{C}|^T\hat{\bm{m}}|\bm{C}|\rfloor)\otimes \textbf{I}_3,
    \label{M}
\end{align}
where $\bm{m} \in \mathbb{R}^{n_e\times 1}$ is the mass vector.

We should point out that in the TTS, the rest length and mass vectors ($\bm{l}_0$ and $\bm{m}$) of the structure members are given constants. However, in the clustered tensegrity, the $\bm{l}_{0}$ and $\bm{m}$ will be changing at each time step ($\bm{l}_{0c}$ and $\bm{m}_c$ are not changing), since the clustered strings are sliding along the pulleys in the dynamics. Thus, the rest length vector $\bm{l}_0$ of all the structure members should be recalculated in every time step by Eqs.(\ref{t_c force vector}) and (\ref{t force vector}), and the mass vector $\bm{m}$ in the CTS is renewed by Eq.(\ref{mass vector}):
\begin{equation}
    \bm{m}=\rho\hat{\bm{A}}(\reallywidehat{\bm{\mathcal{S}}^T\bm{t}_c}+\bm{\hat{E}\hat{A}})^{-1}\bm{\hat{E}\hat{A}}\bm{l}.
\end{equation}


Since the mass matrix $\bm{M}$ is symmetric, we have the following equation:
\begin{align}
    \frac{\mathrm{d}}{\mathrm{d}t}\frac{\partial T}{\partial \dot{\bm{n}}}=\bm{M}\ddot{\bm{n}}.
    \label{Mddn}
\end{align}
Note that we use denominator layout notation in the matrix calculus, which means the derivative of a scalar by a column vector is still a column vector.


\subsubsection{Strain potential energy}
We consider elastic and plastic deformation of structure members. To unify the two cases, the strain potential energy $V_e$ of the whole structure caused by the structure elements' internal force can be written into an integral form:
\begin{equation}
  \label{Ve}
V_e=\sum\limits_i^{n_e}V_{ei}=\sum\limits_i^{n_e}\int_{l_{0i}}^{l_i}t_i\mathrm{d}u,
\end{equation}
where $\mathrm{d}u$ is the differential of the structure member length. The derivative of the strain potential energy $V_e$ with respect to the nodal coordinate vector $\bm{n}$ is:
\begin{equation}
    \frac{\partial V_e}{\partial\bm{n}}
    =\sum\limits_i^{n_e}\frac{\partial V_{ei}}{\partial l_i}\frac{\partial l_i}{\partial \bm{n}}
=\sum\limits_i^{n_e}t_i\frac{\partial l_i}{\partial \bm{n}}.
    \label{dV_edn1}
\end{equation}
The derivative of the structure element's length $l_i$ with respect to the nodal coordinate vector $\bm{n}$ can be obtained from Eq. (\ref{li}):
\begin{align}
    \frac{\partial l_i}{\partial \bm{n}}=\frac{(\bm{C}_i^T\bm{C}_i)\otimes\textbf{I}_3\bm{n}}{l_i}
    \label{dli_dni}.
\end{align}
Substitute Eq. (\ref{dli_dni}) into Eq. (\ref{dV_edn1}), and by the definition of force density $x_i = f_i/l_i$ in the \textit{i}th structure element, we have:
\begin{align}
    \frac{\partial V_e}{\partial\bm{n}}&=\sum\limits_i^{n_e}x_i(\bm{C}_i^T\bm{C}_i)\otimes\textbf{I}_3\bm{n}\\
    &=(\bm{C}^T\reallywidehat{\hat{\bm{l}}^{-1}\bm{\mathcal{S}}^T\bm{t}_c}\bm{C})\otimes\textbf{I}_3\bm{n}\\
    &=\bm{Kn},
    \label{Kn}
\end{align}
where $\bm{K} \in \mathbb{R}^{3n_n \times 3n_n}$ is the stiffness matrix of the tensegrity structure:
\begin{align}
    \bm{K}=(\bm{C}^T\reallywidehat{\hat{\bm{l}}^{-1}\bm{\mathcal{S}}^T\bm{t}_c}\bm{C})\otimes\textbf{I}_3,
    \label{K}
\end{align}
and the force vector $\bm{t}_c$ of the CTS can be calculated by Eq. (\ref{t_c force vector}).

\subsubsection{Gravitational potential energy}

In many cases, the tensegrity structures are in the presence of the gravity field. Let $a_x$, $a_y$, and $a_z$ be the values of the gravity acceleration in the X-, Y-, and Z-direction. The gravitational potential energy $V_g$ can be written as:
\begin{align} 
    V_g &=\sum\limits_i^{n_e}\frac{m_i}{2}    \begin{bmatrix}
    a_x & a_y & a_z
    \end{bmatrix}    \begin{bmatrix}
    x^i_j+x^i_k \\y^i_j+y^i_k \\ z^i_j+z^i_k
    \end{bmatrix}\\ 
    &=\sum\limits_i^{n_e}\frac{m_i}{2} \otimes   \begin{bmatrix}
    a_x & a_y & a_z
    \end{bmatrix}   |\bm{C}_i|\otimes \bm{I}_3\bm{n}\\ 
    &=\frac{1}{2}(\bm{m}^T|\bm{C}|)\otimes    \begin{bmatrix}
      a_x & a_y & a_z
    \end{bmatrix}\bm{n},
\end{align}
where $x^i_j$ and $x^i_k$ are X-coordinates of node $\bm{n}_j$ and node $\bm{n}_k$ of the \textit{i}th structure element. Then, the X-coordinate of the center of mass of the \textit{i}th element is $(x^i_j+x^i_k)/2$. And the partial derivative of $V_g$ with respect to $\bm{n}$ is:
\begin{align}
    \frac{\partial V_g}{\partial \bm{n}}=\frac{1}{2}(|\bm{C}|^T\bm{m})\otimes    \begin{bmatrix}
    a_x & a_y & a_z
    \end{bmatrix}^T=\bm{g},
    \label{g}
\end{align}
where $\bm{g} \in \mathbb{R}^{3 n_n}$ is the gravitational force vector in all the nodes. For structure analysis without gravity, one can set $\bm{g}$ = $\bm{0}$. 

\subsubsection{Damping force}
Damping force is a non-potential force, and here, we directly give the equations to calculate the material damping force. Using the viscous damping constitutive relation, the damping force of all the structure members in the CTS is \cite{kan2018nonlinear}:
\begin{align}
  \bm{f}_{de}&=\hat{\bm{d}}\dot{\bm{l}}_c 
  =\hat{\bm{d}}\frac{\partial\bm{l}_c}{\partial t}
  =\hat{\bm{d}}(\frac{\partial\bm{l}_c}{\partial \bm{n}})^T\frac{\partial\bm{n}}{\partial t}
  =\hat{\bm{d}}\bm{A}_{2c}^T\dot{\bm{n}},
\end{align}
where $\bm{d}$ is the damping coefficient vector of all the structure members, $ \bm{A}_{2c} =\left(\bm{C}^{T} \otimes \textbf{I}_{3}\right) \bm{b.d.}(\bm{H})\bm{\mathcal{S}}^T \hat{\bm{l}}_c^{-1} $ is the equilibrium matrix for clustered tensegrity which is explained in Eq. (\ref{A_{1a}}). 


The damping coefficient should be given by the experiment of the materials. However, since steel and aluminum bars are widely used, here for convenience, we also provide the damping coefficients $\zeta$ of steel and aluminum bars, which and $\zeta_{steel}=1\times10^{-4} \sim 3\times10^{-4}$ and $\zeta_{aluminum}=0.5\times10^{-4}$ \cite{cremer2013structure} of its critical damping coefficient $\bm{d}_{ci}$, where $\bm{d}_{ci}=\frac{2A_{ci}}{3}\sqrt{3\rho_{ci} E_{ci}}$. So the critical damping coefficient vector $\bm{d}_c$ of a structure is suggested to be:
\begin{equation}
    \bm{d}_c=\frac{2\sqrt{3}}{3}\hat{\rho}_{c}^{\frac{1}{2}}\hat{\bm{A}}_c{\bm{E}}_c^{\frac{1}{2}}.
\end{equation}

Distribute the structure members' damping forces to all the nodes, we have the damping force vector:
\begin{align}
  \bm{f}_{d}  &=\zeta\bm{A}_{2c} \bm{f}_{de}
  = \zeta \bm{A}_{2c}\hat{\bm{d}}_c\bm{A}_{2c}^T\dot{\bm{n}}
  =\zeta\bm{D}\dot{\bm{n}},
\end{align}
where $\bm{D}\in \mathbb{R}^{3n_n\times 3n_n}$is the damping matrix:
\begin{equation}
     \bm{D}=\bm{A}_{2c}\hat{\bm{d}}_c\bm{A}_{2c}^T.
     \label{damping matrix D}
\end{equation}

\subsection{Tensegrity dynamics formulation based on Lagrangian method}




For tensegrity structures with boundary constraints, there are $n_a$ number of free nodes as described in Section \ref{Generalized}, the generalized coordinate reduced into $\bm{n}_a$. Then, the Lagrange's equation is: 
\begin{align}\frac{\mathrm{d}}{\mathrm{dt}}\left(\frac{\partial L}{\partial \dot{\bm{n}}_{a}}\right)-\frac{\partial L}{\partial \bm{n}_{a}}=\bm{f}_{n p a},
\label{Lagrange's in na}
\end{align}
where $\bm{f}_{n p a}$ is the non-potential force exerted on free nodal coordinate, and its relation with $\bm{f}_{n p}$ is:
\begin{align}\bm{f}_{n p a}=\bm{E}_{a}^{T} \bm{f}_{n p}.
\label{f_npa}
\end{align}
The non-potential force $\bm{f}_{np}$ is the sum of damping force $\bm{f}_d$ and external force $\bm{f}_{ex}$:
\begin{align}
    \bm{f}_{np}=\bm{f}_{d}+\bm{f}_{ex}.
    \label{f_np}
\end{align}
 The left side of Eq. (\ref{Lagrange's in na}) is obtained by substituting Eqs. (\ref{n=[Ea Eb][na;nb]}), (\ref{L}), (\ref{Mddn}), (\ref{Kn}), (\ref{g}) into Eq. (\ref{Lagarangian's equation}):
\begin{align}
\frac{\mathrm{d}}{\mathrm{dt}}\left(\frac{\partial L}{\partial \dot{\bm{n}}_{a}}\right) &-\frac{\partial L}{\partial \bm{n}_{a}}=\frac{\partial \bm{n}}{\partial \bm{n}_{a}}\left[\frac{\mathrm{d}}{\mathrm{dt}}\left(\frac{\partial L}{\partial \dot{\bm{n}}}\right)-\frac{\partial L}{\partial \bm{n}}\right] \\
&=\bm{E}_{a}^{T}\left[\frac{\mathrm{d}}{\mathrm{dt}}\left(\frac{\partial L}{\partial \dot{\bm{n}}}\right)-\frac{\partial L}{\partial \bm{n}}\right]\\
&=\bm{E}_{a}^{T}(\bm{M} \ddot{\bm{n}}+\bm{K} \bm{n}+\bm{g}).
\label{Lagrange's equ na left}
\end{align}
Substitute Eqs. (\ref{f_npa}), (\ref{f_np}), (\ref{damping matrix D}), (\ref{Lagrange's equ na left}) into Eq. (\ref{Lagrange's in na}), we have the dynamics of clustered tensegrity structures:
\begin{align}\bm{E}_{a}^{T}(\bm{M} \ddot{\bm{n}}+\bm{D} \dot{\bm{n}}+\bm{K} \bm{n})=\bm{E}_{a}^{T}\left(\bm{f}_{e x}-\bm{g}\right).
\label{dynamic in na 1}
\end{align}

Substitute Eq. (\ref{n=[Ea Eb][na;nb]}) into Eq. (\ref{dynamic in na 1}) and rearrange terms related to $\bm{n}_a$, one can obtain:
\begin{align}\nonumber
\bm{M}_{a a} \ddot{\bm{n}}_{a}+\bm{D}_{a a} \dot{\bm{n}}_{a}+\bm{K}_{a a} \bm{n}_{a}= & \bm{E}_{a}^{T} \bm{f}_{e x} -\bm{M}_{a b} \ddot{\bm{n}}_{b}-\bm{D}_{a b} \dot{\bm{n}}_{b} \\ & -\bm{K}_{a b} \bm{n}_{b}-\bm{E}_{a}^{T} \bm{g},
\label{dynamics equation reduced order}
\end{align}
where
\begin{align}\label{aa_ab1}
    \bm{M}_{aa} & =\bm{E}_{a}^{T} \bm{M} \bm{E}_{a},~
\bm{M}_{ab}=\bm{E}_{a}^{T} \bm{M} \bm{E}_{b}, \\ \label{aa_ab2}
\bm{D}_{aa} & =\bm{E}_{a}^{T} \bm{D} \bm{E}_{a},~
\bm{D}_{ab}=\bm{E}_{a}^{T} \bm{D} \bm{E}_{b}, \\ \label{aa_ab3}
\bm{K}_{aa} & =\bm{E}_{a}^{T} \bm{K} \bm{E}_{a},~
\bm{K}_{ab}=\bm{E}_{a}^{T} \bm{K} \bm{E}_{b},
\end{align}
where $\bm{M}$, $\bm{D}$, $\bm{K}$, and $\bm{g}$ are given in Eqs. (\ref{M}), (\ref{damping matrix D}), (\ref{K}), and (\ref{g}). One may also use the following form for programming convenience:
\begin{align}
\ddot{\bm{n}}_{a}=\bm{M}_{a a}^{-1} \bm{E}_{a}^{T}\left(\bm{f}_{e x}-\bm{g}-\bm{M} \bm{E}_{b} \ddot{\bm{n}}_{b}-\bm{D} \dot{\bm{n}}-\bm{K} \bm{n}\right).
\label{n_a_dd}
\end{align}




\section{Statics of clustered tensegrity structures}
\label{section4}

In this section, we give the equilibrium equations in three standard forms and the compatibility equation. The equations developed in this section are useful for statics analysis and the derivation of the linearized dynamics in the next section. \par 

\subsection{Equilibrium equations}

Let the acceleration part $\ddot{\bm{n}}$ and velocity part $\dot{\bm{n}}$ in Eq. (\ref{dynamic in na 1})  be zeros, the dynamics equation will be reduced into the static equilibrium equation in terms of nodal coordinate vector $\bm{n}$ or free nodal coordinate $\bm{n_a}$:
\begin{align}\label{static equilibrium1}
\bm{E}_{a}^{T}\bm{K} \bm{n}=\bm{E}_{a}^{T}(\bm{f}_{e x}-\bm{g}).
\end{align}
Similarly, we have an equivalent form obtained from Eq. (\ref{dynamics equation reduced order}):
\begin{align}
\bm{K}_{a a} \bm{n}_{a}= & \bm{E}_{a}^{T}(\bm{f}_{e x}-\bm{g}) -\bm{K}_{a b} \bm{n}_{b}.
\label{static equilibrium reduced order}
\end{align}

Since $\bm{K}$, given in Eq. (\ref{K}), is a function of $\bm{n}$, the product $\bm{Kn}$ is nonlinear in $\bm{n}$. Eq. (\ref{static equilibrium1}) is a nonlinear equilibrium equation. However, the term $\bm{Kn}$ can be written linearly in terms of force vector $\bm{t}_c$:
\begin{align}
\bm{K} \bm{n} &=\left(\bm{C}^{T} \otimes \textbf{I}_{3}\right)(\reallywidehat{\hat{\bm{l}}^{-1}\mathcal{S}^T\bm{t}_c} \otimes \textbf{I}_{3})\left(\bm{C} \otimes \textbf{I}_{3}\right) \bm{n} \\
&=\left(\bm{C}^{T} \otimes \textbf{I}_{3}\right) \reallywidehat{\left(\textbf{I}_{n_{e}} \otimes \textbf{I}_{3,1} \hat{\bm{l}}^{-1}\mathcal{S}^T\bm{t}_c\right)}\left(\bm{C} \otimes \textbf{I}_{3}\right) \bm{n} \\
&=\left(\bm{C}^{T} \otimes \textbf{I}_{3}\right)\reallywidehat{\left(\left(\bm{C} \otimes \textbf{I}_{3}\right) \bm{n}\right)} \textbf{I}_{n_{e}} \otimes \textbf{I}_{3,1} \hat{\bm{l}}^{-1}\bm{\mathcal{S}}^T\bm{t}_c\\
&=\left(\bm{C}^{T} \otimes \textbf{I}_{3}\right) \bm{b.d.}(\bm{H}) \hat{\bm{l}}^{-1}\bm{\mathcal{S}}^T\bm{t}_c
\label{Kn=~~~x}
\end{align}
where $\bm{b.d.}(V)$ is the block diagonal matrix of $V$. Substitute Eq. (\ref{Kn=~~~x}) into Eq. (\ref{static equilibrium1}), the equilibrium equation can be written linearly in terms of force vector $\bm{t}_c$:
\begin{align}
\bm{E}_{a}^{T}\bm{A}_{2c}\bm{t}_c=\bm{E}_{a}^{T}(\bm{f}_{e x}-\bm{g}),
\label{static equilibrium3}
\end{align}
where
\begin{align}
\bm{A}_{2c}=\left(\bm{C}^{T} \otimes \textbf{I}_{3}\right) \bm{b.d.}(\bm{H}) \hat{\bm{l}}^{-1}\bm{\mathcal{S}}^T.
\label{A_{1a}}
\end{align}
 \par
 

The equilibrium equation can also be written linearly in terms of force density vector $\bm{x}_c$ by substitute Eq. (\ref{x_bar force density}) into Eq. (\ref{static equilibrium3}):
\begin{align}
\bm{E}_{a}^{T}\bm{A}_{1c} \bm{x}_c=\bm{E}_{a}^{T}(\bm{f}_{e x}-\bm{g}),
\label{static equilibrium2}
\end{align}
where
\begin{align}
\bm{A}_{1c} =\bm{A}_{2c} \hat{\bm{l}}_c =\left(\bm{C}^{T} \otimes \textbf{I}_{3}\right) \bm{b.d.}(\bm{H}) \hat{\bm{l}}^{-1}\bm{\mathcal{S}}^T \hat{\bm{l}}_c.
\label{equilibrium_A2c}
\end{align}

If the structure is in a self-equilibrium state and there are no constraints, the equilibrium equation can be obtained by setting $\bm{E_a}$ equals identity matrix in Eqs. (\ref{Kn=~~~x}), (\ref{static equilibrium2}) and (\ref{static equilibrium3}).

\subsection{Compatibility equations}

Compatibility equation is the relation between nodal coordinate $\bm{n}$ and structure member length $\bm{l}$. Stack Eq. (\ref{dli_dni}) in a column, one can obtain the compatibility equation:
\begin{align}\bm{B}_{l}  \mathrm{d} \bm{n}= \mathrm{d} \bm{l},
\label{compatibility equation}
\end{align}
where $\bm{B}_{l}\in \mathbb{R}^{n_e\times 3 n_n }$ is the compatibility matrix of the TTS:
\begin{align}\bm{B}_{l}=\hat{\bm{l}}^{-1} \bm{b.d.}(\bm{H})^{T}\left(\bm{C} \otimes \textbf{I}_{3}\right).
\label{Bl}
\end{align}
For the CTS, the compatibility equation is obtained by substituting Eq.(\ref{compatibility equation}) into the differential of Eq.(\ref{l_c=Sl}):
\begin{equation}
    \bm{B}_{lc}  \mathrm{d} \bm{n}= \mathrm{d} \bm{l}_c,
    \label{B_lc dn=dl_c}
\end{equation}
where 
$\bm{B}_{lc}\in \mathbb{R}^{n_{ec}\times 3 n_n }$ is the compatibility matrix of the CTS:
\begin{align}\bm{B}_{lc}=\bm{\mathcal{S}}\hat{\bm{l}}^{-1} \bm{b.d.}(\bm{H})^{T}\left(\bm{C} \otimes \textbf{I}_{3}\right).
\label{B_lc}
\end{align}

Note that the compatibility and equilibrium matrix of the CTS has the following relationship: $\bm{B}_{lc}^{T}=\bm{A}_{2c}$.


\section{Linearized tensegrity dynamics}
\label{section5}


From Eq. (\ref{t_c force vector}), we know the structure equilibrium is influenced by the nodal coordinate $\bm{n}$ and the rest length of members $\bm{l}_{0c}$ if the cross-sectional area, Young's modulus of the structure members are constant. Take the total derivative of Eq. (\ref{dynamic in na 1}) and keep the linear terms, one can have the linearized dynamics:
\begin{align}\nonumber
&\bm{E}_{a}^{T}
\left[\frac{\partial(\bm{M\ddot{\bm{n}}})}{\partial\ddot{\bm{n}}}\right]^T\mathrm{d} \ddot{\bm{n}}+\left[\frac{\partial(\bm{D\dot{\bm{n}}})}{\partial\dot{\bm{n}}}\right]^T\mathrm{d} \dot{\bm{n}}
+\left[\frac{\partial(\bm{K{\bm{n}}})}{\partial{\bm{n}}}\right]^T\mathrm{d} {\bm{n}}\\&+\left[\frac{\partial(\bm{K{\bm{n}}})}{\partial{\bm{l}}_{0c}}\right]^T\mathrm{d} {\bm{l}_{0c}}
=\bm{E}_{a}^{T}\mathrm{d}\bm{f}_{e x},
\label{linear dynamics with constraints 1}
\end{align}
which is equivalent to:
\begin{align}
\bm{E}_{a}^{T}(\bm{M}\mathrm{d} \ddot{\bm{n}}+\bm{D} \mathrm{d}\dot{\bm{n}}+\bm{K}_T\mathrm{d} \bm{n}+\bm{K}_{\bm{l}_{0c}}\mathrm{d} \bm{l}_{0c})=\bm{E}_{a}^{T}\mathrm{d}\bm{f}_{e x},
\label{linear dynamics with constraints}
\end{align}
where $\bm{K}_T$ is the tangent stiffness matrix given in Eq. (\ref{Kt tangent stiffness matrix 2}) and $\bm{K}_{{\bm{l}}_{0c}}$ is the sensitivity matrix of the rest length to the nodal force, given in Eq. (\ref{K_l0c}). And Eq. (\ref{linear dynamics with constraints}) has the following equivalent form where the free and constrained node vectors are separated on the two sides of an equation: 
\begin{align}\nonumber
& \bm{M}_{a a} \mathrm{d}\ddot{\bm{n}}_{a} + \bm{D}_{a a} \mathrm{d}\dot{\bm{n}}_{a}+\bm{K}_{Ta a} \mathrm{d}\bm{n}_{a}\\ \nonumber
= & \bm{E}_{a}^{T} \mathrm{d} \bm{f}_{e x}-\bm{E}_{a}^{T} \bm{K}_{\bm{l}_{0c}} \mathrm{d} \bm{l}_{0c} -\bm{M}_{a b} \mathrm{d}\ddot{\bm{n}}_{b} \\  \label{reduced order dynamic}
& -\bm{D}_{a b} \mathrm{d}\dot{\bm{n}}_{b} -\bm{K}_{Tab} \mathrm{d}\bm{n}_{b},
\end{align}
where:
\begin{align}
 \bm{K}_{Taa} & =\bm{E}_{a}^{T} \bm{K}_T \bm{E}_{a},~\bm{K}_{Tab} =\bm{E}_{a}^{T} \bm{K}_T \bm{E}_{b},
\end{align}
and $\bm{M}_{a a}$, $\bm{M}_{a b}$, $\bm{D}_{a a}$, $\bm{D}_{a b}$ are given in Eqs. (\ref{aa_ab1}) - (\ref{aa_ab3}).
One can write the linearized dynamics equation with constraints into a state-space form:
\begin{equation}
\begin{aligned}
\frac{\mathrm{d}}{\mathrm{d}t} \begin{bmatrix}\mathrm{d}\bm{n}_a\\\mathrm{d}\dot{\bm{n}}_{a}\end{bmatrix}&=\begin{bmatrix}\bm{ 0}  & \textbf{I}\\ -\bm{M}_{aa}^{-1}K_{Ta a} & \bm{-M}_{aa}^{-1} \bm{D}_{aa}\end{bmatrix}    \begin{bmatrix}\mathrm{d}\bm{n}_a\\\mathrm{d}\dot{\bm{n}}_a\end{bmatrix}\\&+ \begin{bmatrix}\bm{0}&\bm{0}\\ \bm{M}_{aa}^{-1}\bm{E}_{a}^{T} &- \bm{M}_{aa}^{-1}\bm{E}_{a}^{T}\bm{K}_{l_0 a} \end{bmatrix}\begin{bmatrix} \mathrm{d}\bm{f}_{ex} \\ \mathrm{d}\bm{l}_{0c} \end{bmatrix} \\
 & + \begin{bmatrix} \bm{0} \\ \bm{M}_{aa}^{-1} (
-\bm{M}_{a b} \mathrm{d}\ddot{\bm{n}}_{b}-\bm{D}_{a b} \mathrm{d}\dot{\bm{n}}_{b} -\bm{K}_{T ab} \mathrm{d}\bm{n}_{b})
\end{bmatrix},
\end{aligned}
\label{dynamic equation with constraints in first order}
\end{equation}
which can be an interface to integrate structure and control designs.

\subsection{Tangent stiffness matrix}
The tangent stiffness matrix $\bm{K}_T$ is calculated as:
\begin{align}
\bm{K}_{T}=\left[\frac{\partial(\bm{K} \bm{n})}{\partial \bm{n}}\right]^{T}=\bm{K}+\left[\frac{\partial \bm{t}_c}{\partial \bm{n}} \frac{\partial(\bm{K} \bm{n})}{\partial \bm{t}_c}\right]^{T}.
\label{Kt tangent stiffness matrix 0}
\end{align}
The partial derivative of force vector $\bm{t}_c$ with respect to the nodal coordinate vector $\bm{n}$ can be obtained from Eqs. (\ref{t_c force vector}) and (\ref{B_lc dn=dl_c}):
\begin{align}
\frac{\partial \bm{t}_c}{\partial \bm{n}} &=\frac{\partial\left[\hat{\bm{E}}_{tc}\hat{\bm{A}}_c\hat{\bm{l}}_{0c}^{-1}(\bm{l}_c-\bm{l}_{0c})\right]}{\partial \bm{n}} \\
&=\bm{A}_{2c} \hat{\bm{l}}_{0c}^{-1} \widehat{\bm{A}}_c \widehat{\bm{E}}_{tc}.
\label{x by n}
\end{align}

The derivative of $\bm{Kn}$ with respect to force vector $\bm{t}_c$ is derived from Eq. (\ref{Kn=~~~x}), then we have: 
\begin{align}\frac{\partial(\bm{K} \bm{n})}{\partial \bm{t}_c}=\frac{\partial\left(\bm{A}_{2c} \bm{t}_c\right)}{\partial \bm{t}_c}=\bm{A}_{2c}^{T}.
\label{kn by x}
\end{align}
Substitute Eqs. (\ref{x by n}) and (\ref{kn by x}) into Eq. (\ref{Kt tangent stiffness matrix 0}), one can obtain the tangent stiffness matrix $\bm{K}_{T}$:
\begin{align}
\bm{K}_{T}=\left(\bm{C}^{T} \widehat{\hat{\bm{l}}^{-1}\bm{\mathcal{S}}^T\bm{t}_c} \bm{C}\right) \otimes \textbf{I}_{3}+\bm{A}_{2c} \widehat{\bm{E}}_{tc} \widehat{\bm{A}}_c \hat{\bm{l}}_{0c}^{-1} \bm{A}_{2c}^{T}.
\label{Kt tangent stiffness matrix 2}
\end{align}
The first part of Eq. (\ref{Kt tangent stiffness matrix 2}) is usually called the geometry stiffness matrix $\bm{K}_{G}=\left(\bm{C}^{T} \widehat{\hat{\bm{l}}^{-1}\bm{\mathcal{S}}^T\bm{t}_c} \bm{C}\right) \otimes \textbf{I}_{3}$, which is determined by structure topology and force density. The second part is called the material stiffness $\bm{K}_{E}=\bm{A}_{2c} \widehat{\bm{E}}_{tc} \widehat{\bm{A}}_c \hat{\bm{l}}_{0c}^{-1} \bm{A}_{2c}^{T}$, which is governed by structure configuration and structure elements' axial stiffness. By setting $\bm{\mathcal{S}=I}$ in Eq. (\ref{Kt tangent stiffness matrix 2}), the tangent stiffness matrix of CTS is the same as TTS, the result is also consistent with \cite{zhang2006adaptive}.

\subsection{Sensitivity matrix}
The sensitivity matrix represent the sensitivity of changes in the structure members rest length to changes in structure's nodal force: 
\begin{align}
   \bm{K}_{\bm{l}_{0c}}&= \left[\frac{\partial(\bm{K{\bm{n}}})}{\partial{\bm{l}}_{0c}}\right]^T\\
   &=\left[\frac{\partial \bm{x}_c}{\partial\bm{l}_{0c}}\frac{\partial\left(\bm{A}_{1c} \bm{x}_c\right)}{\partial \bm{x}_c}   \right]^T\\
   &=-\bm{A}_{1c}\widehat{\bm{E}}_{tc} \widehat{\bm{A}}_c \hat{\bm{l}}_{0c}^{-2}\label{K_l0c}.
\end{align}

\subsection{Modal analysis of the linearized model with constraints}
For tensegrity dynamics with constraints, the free vibration response can be obtained from Eq. (\ref{reduced order dynamic}) by neglecting damping force, external force, change of rest length and motion of boundary nodes:
\begin{align}
\bm{M}_{aa}\mathrm{d} \ddot{\bm{n}}_a+\bm{K}_{Taa}\mathrm{d} \bm{n}_a=\bm{0}.
\label{free vibration with constraints}
\end{align}
The solution to the homogeneous Eq.  (\ref{free vibration with constraints}) have the following form:
\begin{align} \mathrm{d} \bm{n}_a=\bm{\varphi} \sin (\omega t-\theta),
\label{dn_a}\end{align}
which represents a periodic response with a frequency $\omega$. Substitute Eq. (\ref{dn_a}) into Eq. (\ref{free vibration with constraints}), we have:
\begin{align}
\left(\bm{K}_{T}-\omega^{2} \bm{M}\right) \bm{\varphi} \sin (\omega t-\theta)=\bm{0},
\end{align}
and since $\sin (\omega t-\theta)\ne0$ for most times, we have the generalized eigenvalue problem:
\begin{align}
\bm{K}_{Taa} \bm{\varphi}=\omega^{2}\bm{M}_{aa} \bm{\varphi},
\label{eigenvalue problem with constratints}
\end{align}
where $\omega$ is the natural frequency of the system and $\bm{\varphi}$ is the corresponding eigenvector representing the mode shapes.

\section{Shape control of clustered tensegrity systems}
\label{section6}


Tensegrity has the advantage in deployability due to the abundant strings. However, a general control law is needed to achieve the morphing objectives. In this section, we present a nonlinear control law that is applicable to the shape control of any tensegrity structures (both the TTS and CTS). 

\subsection{Shape objectives}

For any control problem, it is critical to define the nodes of interest and their morphing objectives. Thus, let us first define a matrix $\bm{E}_c$ to extract the nodal coordinate of interest $\bm{n}_c$ from the free nodal coordinate $\bm{n}_a$:
\begin{equation}
    \bm{n}_c =  \bm{E}_c^T \bm{n}_a .\label{nc_bar}
\end{equation}
Then, the morphing objective of $\bm{n}_c$ is noted as $\bm{\bar{n}}_c$, and the time derivatives of $\bm{\bar{n}}_c$ gives the velocity $\dot{\bm{\bar{n}}}_c$ and acceleration $\ddot{\bm{\bar{n}}}_c$. We use a vector $\bm{e}$ to compute the errors between the current position and target  coordinates:
\begin{align}
    \bm{e} = \bm{n}_c - \bm{\bar{n}}_c = \bm{E}_c^T\bm{n}_a - \bm{\bar{n}}_c.
    \label{error e}
\end{align}

\subsection{Active and passive members}

Due to the abundant number of structure members in the tensegrity structure and to reduce the number of actuators for engineering applications, some structure members can be chosen as active members, while others are passive ones. The active structure members can actively change their length, and passive ones move only passively. It is worthy of mentioning that the active members can be both bars (i.e., linear motors, telescopic bars) and strings (i.e., motor driving cables, shape memory alloys). 

To distinguish the active and passive structure members, we use two matrices $\bm{E}_{act}$ and $\bm{E}_{pas}$ to separate all the structure members. The force vectors of clustered tensegrity of the active and passive members can be written as:
\begin{align}
\bm{t}_{c_{act}}=\bm{E}_{act}^{T} \bm{t}_{c},~ \bm{t}_{c_{pas}}=\bm{E}_{pas}^{T}  \bm{t}_{c},
\label{t_c_act}
\end{align}
where $\bm{t}_{c_{act}}$ and $\bm{t}_{c_{pas}}$ are the force vectors of active and passive members respectively. Since $\begin{bmatrix}\bm{E}_{act} &  \bm{E}_{pas}\end{bmatrix}$ is an orthogonal matrix, we have the following equation:
\begin{align}\bm{t}_c=
\begin{bmatrix}
\bm{E}_{act}^{T} \\
\bm{E}_{pas}^{T}
\end{bmatrix}^{-1}
\begin{bmatrix}\bm{t}_{c_{act}} \\
\bm{t}_{c_{pas}}
\end{bmatrix}=
\begin{bmatrix}
\bm{E}_{act} & \bm{E}_{pas}\end{bmatrix}
\begin{bmatrix}
\bm{t}_{c_{act} }\\
\bm{t}_{c_{pas}}
\end{bmatrix}.
\label{t_c=E_act E_pas t_act t_pas}
\end{align}

\subsection{Error dynamics}

To achieve the desired shape control, the error vector $\bm{e}$ and its time derivatives should all go to zero when the nodes of interest reaches their targets. This goal can be expressed as follows:
\begin{align}
    \ddot{\bm{e}} + \bm{\psi} \dot{\bm{e}} + \bm{\phi} \bm{e} = \bm{0},\label{error dynamic of e}
\end{align}
where $\bm{\psi}$ and $\bm{\phi}$ are tune matrices that can adjust the time response of the morphing process. Since $\bm{E}_c$ is given constants, the time derivatives of error vector in Eq. (\ref{error e}) are:
\begin{align}
    \dot{\bm{e}} &= \dot{\bm{n}}_c - \dot{\bm{\bar{n}}}_c,\\
    \ddot{\bm{e}} &= \ddot{\bm{n}}_c - \ddot{\bm{\bar{n}}}_c.
\end{align}
Substitute $\ddot{\bm{e}}$ and $\dot{\bm{e}}$ into Eq. (\ref{error dynamic of e}), we have:
\begin{align}
    (\ddot{\bm{n}}_c - \ddot{\bm{\bar{n}}}_c)  + \bm{\psi} (\dot{\bm{n}}_c - \dot{\bm{\bar{n}}}_c) + \bm{\phi} (\bm{n}_c - \bm{\bar{n}}_c) = \bm{0}. 
\end{align}
Substitute Eq. (\ref{nc_bar}) and its derivative into the above equation, the error dynamics equation can be written with free nodal coordinate:
\begin{align}
    \bm{E}_c^T\ddot{\bm{n}}_a - \ddot{\bm{\bar{n}}}_c + \bm{\psi} (\bm{E}_c^T\dot{\bm{n}}_a - \dot{\bm{\bar{n}}}_c) + \bm{\phi} (\bm{E}_c^T\bm{n}_a- \bm{\bar{n}}_c) = \bm{0}. 
\end{align}
Substitute Eqs. (\ref{n_a_dd}) and (\ref{error e}) into the above equation we have:
\begin{align}\nonumber
& \bm{E}_c^T\bm{M}_{a a}^{-1} \bm{E}_{a}^{T}(\bm{f}_{e x}-\bm{g}-\bm{M} \bm{E}_{b} \ddot{\bm{n}}_{b}-\bm{D} \dot{\bm{n}}-\bm{K} \bm{n}) \\ 
& -  \ddot{\bm{\bar{n}}}_c  + \bm{\psi} (\bm{E}_c^T\dot{\bm{n}}_a - \dot{\bm{\bar{n}}}_c) + \bm{\phi} (\bm{E}_c^T\bm{n}_a- \bm{\bar{n}}_c) = \bm{0}.\label{error dynamics3}
\end{align}

\subsection{Solving for control variable}


Substitute Eqs. (\ref{static equilibrium3}) and (\ref{t_c=E_act E_pas t_act t_pas}) into Eq. (\ref{error dynamics3}), one can have a linear algebra equation:
\begin{align}
    \bm{\mu}-\bm{\Gamma}_{pas} \bm{t}_{c_{pas}} &= \bm{\Gamma}_{act} \bm{t}_{c_{act}},
    \label{mu=gamma t_c}
    \end{align}
where $\bm{\mu}$, $\bm{\Gamma}_{act}$, and $\bm{\Gamma}_{pas}$ are:
\begin{align}\nonumber
    \bm{\mu} &=\bm{E}_c^T\bm{M}_{a a}^{-1} \bm{E}_{a}^{T}(\bm{f}_{e x}-\bm{g}-\bm{M} \bm{E}_{b} \ddot{\bm{n}}_{b}-\bm{D} \dot{\bm{n}}) \\  & -  \ddot{\bm{\bar{n}}}_c  + \bm{\psi} (\bm{E}_c^T\dot{\bm{n}}_a - \dot{\bm{\bar{n}}}_c) + \bm{\phi} (\bm{E}_c^T\bm{n}_a- \bm{\bar{n}}_c), \\
    \bm{\Gamma}_{act}&=E_c^T \bm{M}_{a a}^{-1} \bm{E}_{a}^{T}\bm{A}_{2c}\bm{E}_{act},\\
    \bm{\Gamma}_{pas}&=E_c^T \bm{M}_{a a}^{-1} \bm{E}_{a}^{T}\bm{A}_{2c}\bm{E}_{pas}.
\end{align}




\begin{figure}
    \centering
    \includegraphics[scale=0.35]{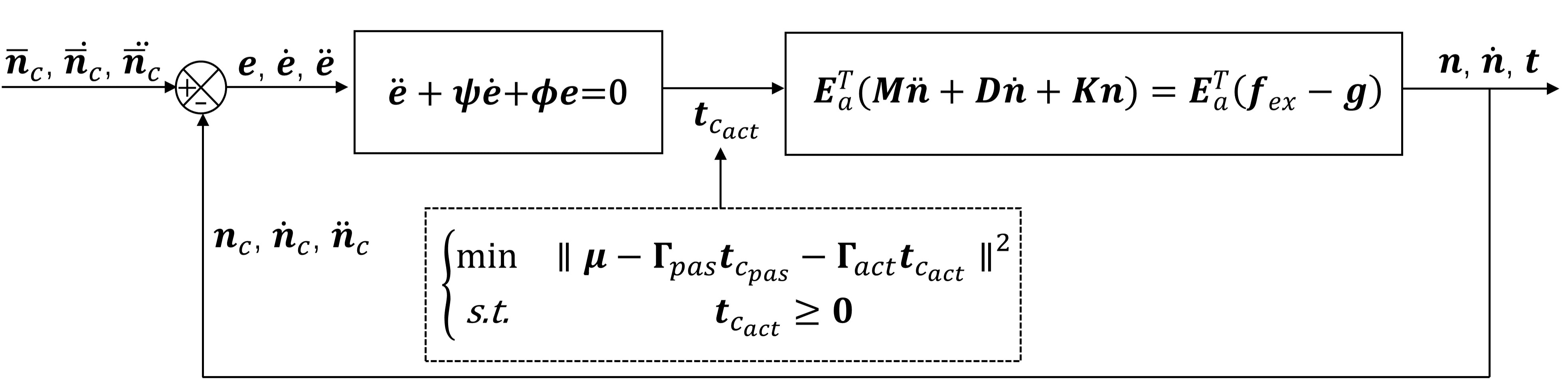}
    \caption{Closed loop control block.}
    \label{control_block}
\end{figure}
The only unknown variable in Eq. (\ref{mu=gamma t_c}) is $\bm{t}_{c_{act}}$. In most cases, people use strings to control the tensegrity structure, so let us use string as the control variable as an example. Since strings cannot take compression, and if they do, we should substitute the force to zero, we must add the constraint to this unknown variable $\bm{t}_{c_{act}} \geq \bm{0}$. Thus, the problem of solving  Eq. (\ref{mu=gamma t_c}) becomes a linear algebra problem with inequality constraints. The solution depends on the properties of $\bm{\Gamma}_{act}$, and for different structures, one cannot always guarantee a solution. One can use a least square formulation to solve the problem at each increment of real time $\Delta t$ is:
\begin{align}
\begin{cases}
    \min  &||\bm{\mu}-\bm{\Gamma}_{pas} \bm{t}_{c_{pas}} - \bm{\Gamma}_{act} \bm{t}_{c_{act}}||^2 \\
    s.t. & \bm{t}_{c_{act}}\geq \bm{0}
    \end{cases}. \label{lsqin}
\end{align}
The solution to the above equation provides the member force needed for shape control in each time step. For some other applications, one may want to change the rest length of the strings, and the rest length of active members can be calculated by Eq. (\ref{t_c force vector}):
\begin{align}
    \bm{l}_{0c_{act}}=(\hat{\bm{t}}_{c_{act}}+\hat{\bm{E}}_{c_{act}}\hat{\bm{A}}_{c_{act}})^{-1}\hat{\bm{E}}_{c_{act}}\hat{\bm{A}}_{c_{act}}\bm{l}_{c_{act}}.
\end{align}





\section{Numerical examples}
\label{section7}




In this section, three representative numerical examples are investigated to prove the accuracy and efficiency of the proposed static, dynamic, and control theories of the CTS.

\subsection{A planer clustered T-bar}

We start with a 2D T-Bar structure, which has been proved to be a mass efficient structure \cite{skelton2016globally,skelton2009tensegrity}, to demonstrate the prestress design, quasi-static statics analysis, dynamics, and successful control of it. 

\begin{figure}
    \centering
    \includegraphics[scale=0.45]{ 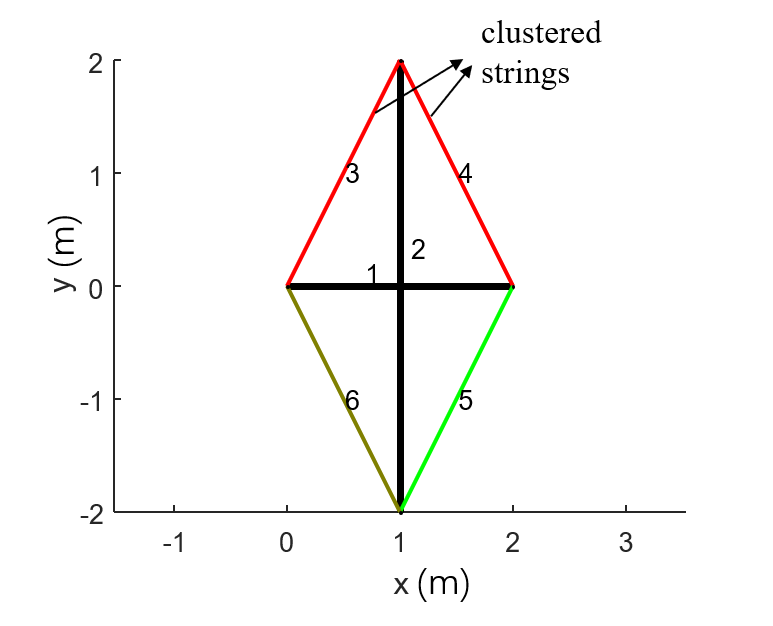}
    \caption{A planar clustered T-bar.}
    \label{planer clustered T-bar}
\end{figure}

A classical T-bar structure is composed of four separate strings, and two bars \cite{skelton2009tensegrity}. We transform it into a cluster one by connecting the members 3 and 4 into one clustered string, as shown in Fig.\ref{planer clustered T-bar}. Note that bars are in black, strings are in other colors, and the clustered strings are in the same color. \par


\begin{table}[]
    \centering
    \begin{tabular}{ll}
         \hline
         Parameter & Values\\
         \hline
         Cross sectional area of strings & $9.138\times10^{-7}$ ${\rm m}^2$\\
         Cross sectional area of horizontal bar & $1.57\times10^{-4}$ ${\rm m}^2$\\
         Cross sectional area of vertical bar & $4.447\times10^{-4}$ ${\rm m}^2$\\
         Young's modulus of bars & $2.06\times10^{11}$ {\rm Pa}\\ 
         Young's modulus of strings & $7.6\times10^{10}$ {\rm Pa}\\
         Density of bars & $7,870$ ${\rm kg}/{\rm m}^3$\\
         Density of strings & $7,870$ ${\rm kg}/{\rm m}^3$\\
         Yielding stress of bars & $435\times10^{6}$ {\rm Pa}\\ 
         Yielding stress of strings & $1,223.5\times10^{6}$ {\rm Pa}\\
         \hline
    \end{tabular}
    \caption{The material parameters of the structure}
    \label{the material parameters of the structure}
\end{table}

\subsubsection{Static analysis}

To keep the structure in a stable equilibrium state, in this example, we first assign the force of bar-1 to be 100N in compression, by the equilibrium matrix Eq. (\ref{equilibrium_A2c}), we can get the forces in all the strings are 200N in tension, and force in bar-2 is 111.8N in compression. The cross-sectional area of members is calculated by 10\% of yielding or buckling load. As shown in Fig.\ref{planer clustered T-bar}, the length of bar-1 and bar-2 are 1m and 2m. The OD (outer diameter) of the bar-1, bar-2, and strings are 14.15mm, 23.79mm, and 1.08mm, respectively.  We use steel cables and Q235 steel rods for strings and bars. The material properties are listed in Table \ref{the material parameters of the structure}. 

\begin{figure}
    \centering
    \includegraphics[scale=0.6]{ 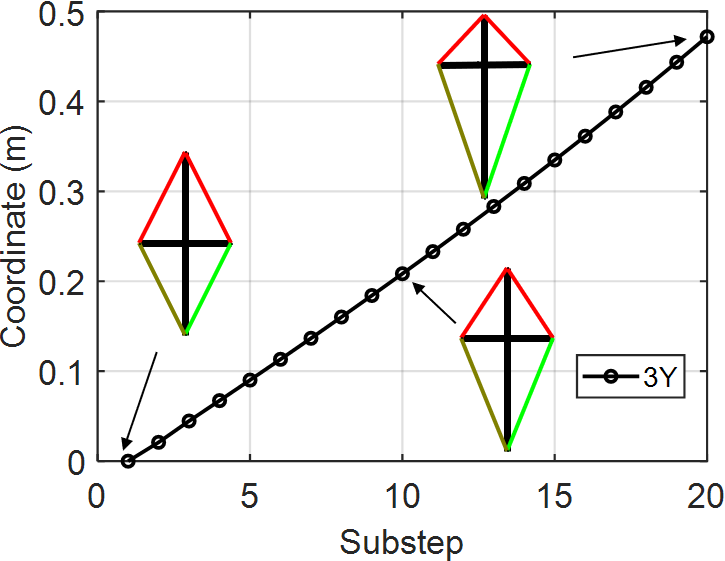}
    \caption{Structure configuration with respect to the substep.}
    \label{structure configuration to substep}
\end{figure}

Here, to demonstrate the nonlinear static analysis in equilibrium state finding, we perform a quasi-static statics analysis. And the results we get here will be compared with the dynamics given in the next section. Firstly, we actively decrease the rest length of the clustered string (a single string replaces strings 3 and 4) by 2m, while the rest length of strings 5 and 6 increasing by 0.5m as the actuation strategy. This actuation process is equally divided into 20 substeps in our program. The structure configuration with respect to actuation substeps is show in Fig.\ref{structure configuration to substep}.  We can see that the structure changes its shape to the desired shape.

\subsubsection{Dynamic analysis}

\begin{figure}
    \centering
    \includegraphics[scale=0.6]{ 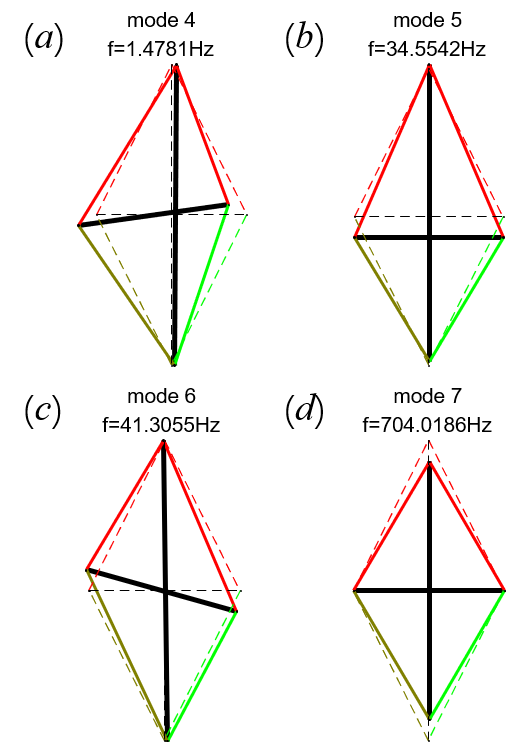}
    \caption{Free vibration modes.}
    \label{free vibration mode}
\end{figure}

For the analysis of the dynamics, we first check the modal analysis. Results show that the first three modes are rigid body motions with zero natural frequencies. The 4th to 7th natural frequencies and vibration modes are shown in Fig. \ref{free vibration mode}. One can see that the 4th vibration mode demonstrates a movement of the pulley in the clustered string. And the frequency 1.4781Hz is relatively much lower compared with the 5th and 6th ones. The result indicates that the use of pulleys and clustered strings will decrease the stiffness of the structure, which agrees with the physics.

We also investigated the nonlinear dynamics of this clustered T-Bar structure. The actuation strategy is the same as the quasi-static one but with various string stretch speeds. The actuation speed is given as follows, in the first half of total time (T=1s, 0.5s, 0.1s, 0.05s), we actuate the strings, while in the second half, we stop the actuation and leave the system free response. The time step is 0.0001s, and the damping coefficient of all the structural materials is 0.01.


\begin{figure}
    \centering
    \includegraphics[scale=0.38]{ 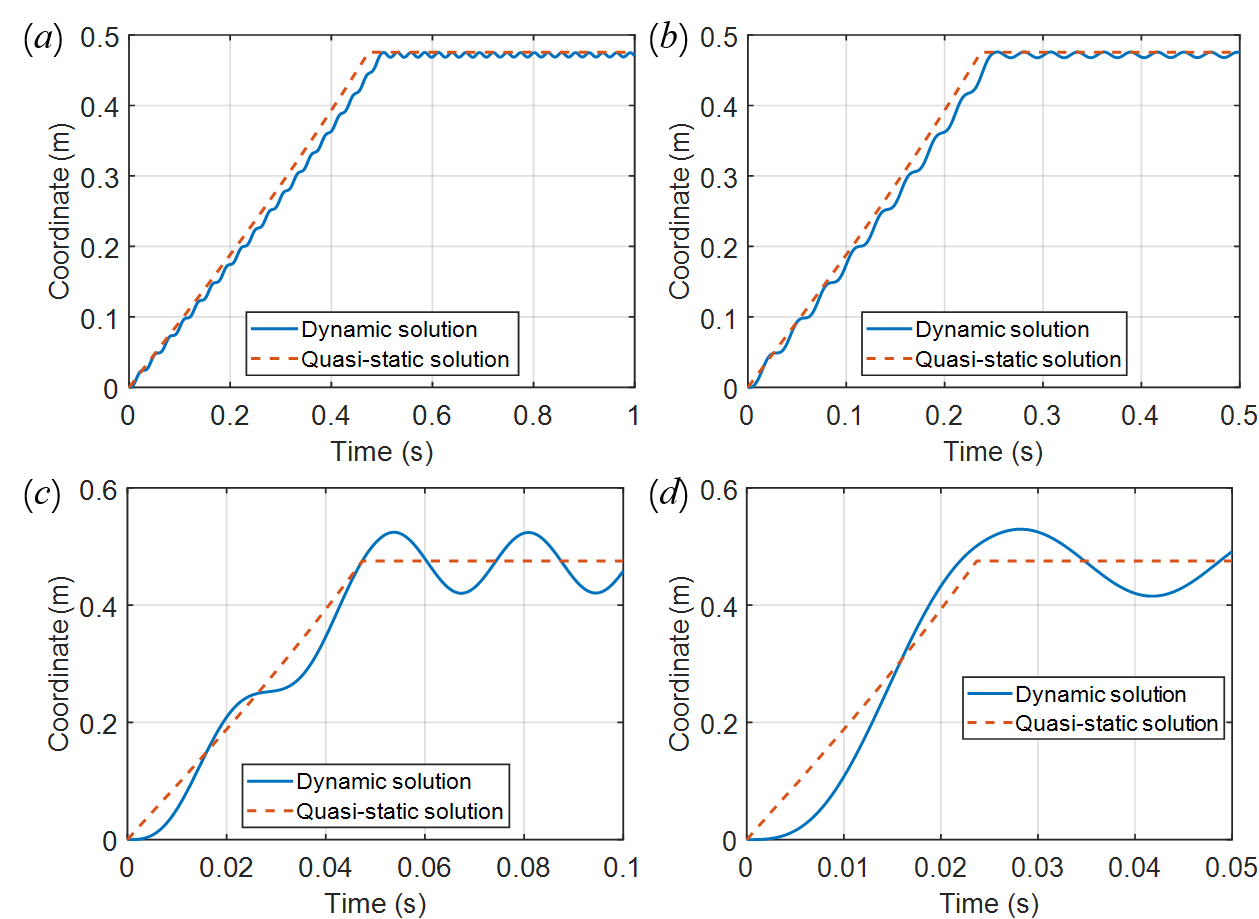}
    \caption{Time history curves of the Y-coordinate of horizontal bar.}
    \label{curves of the Y coordinate of horizontal bar}
\end{figure}

The time history of the Y-coordinate of bar-1 is given by 
Fig.\ref{curves of the Y coordinate of horizontal bar} colored in solid blue. The quasi-static results (dotted red lines) are presented as a comparison. One can observe that the actuation speeds are critical to the dynamic responses. For relatively low actuation speeds (T=1s, 0.5s), the dynamic responses are close to the quasi-static result. The amplitude of vibration becomes larger as the actuation speed increases. And for high actuation speeds (T=0.1s, 0.05s), the dynamic responses shift away from the quasi-static solution.


\subsubsection{Shape control}

From the statics and dynamics analysis of changing the rest length of strings, the structure morphs shape as we expected. However, for engineering applications, the dynamics are always involved, and we would like to have a closed-loop smooth control. This requirement can be achieved by the developed approach in Section \ref{section6}. \par

\begin{figure}
    \centering
    \includegraphics[scale=0.5]{ 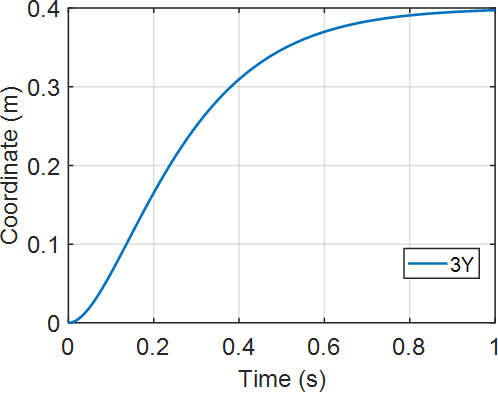}
    \caption{Time history curves of the Y-coordinate of node 3.}
    \label{curves of the y coordinate}
\end{figure}

\begin{figure}
    \centering
    \includegraphics[scale=0.5]{ 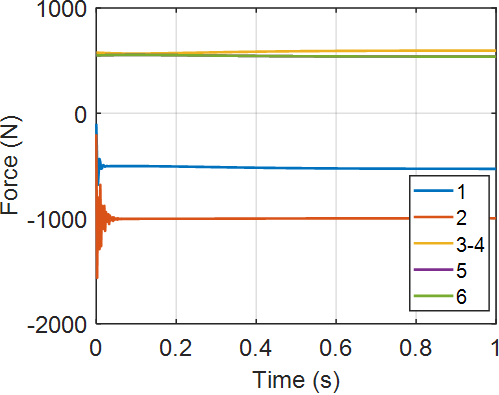}
    \caption{Time history curves of the member force.}
    \label{curves of the member force}
\end{figure}

The shape control objective is set to move the Y-coordinates of nodes 1 and 3 to 0.4m, and the active member is chosen as all the strings. The coefficients in the control law is $\bm{\psi}=2\sqrt{50} \bm{I}$ and $\bm{\phi}=50 \bm{I}$. We should point out that one is free to adjust the two coefficients based on the bandwidth of their actuators and deploy speed needs. The control result is shown in Fig.\ref{curves of the y coordinate}. We can see that the target coordinate is driven from 0 to 0.4 smoothly with no oscillation compared to the nonlinear dynamic simulation. Fig.\ref{curves of the member force} gives the required control forces of each structure member. 

\subsection{A two-stage clustered tensegrity tower}

\begin{figure}
    \centering
    \includegraphics[scale=0.43]{ 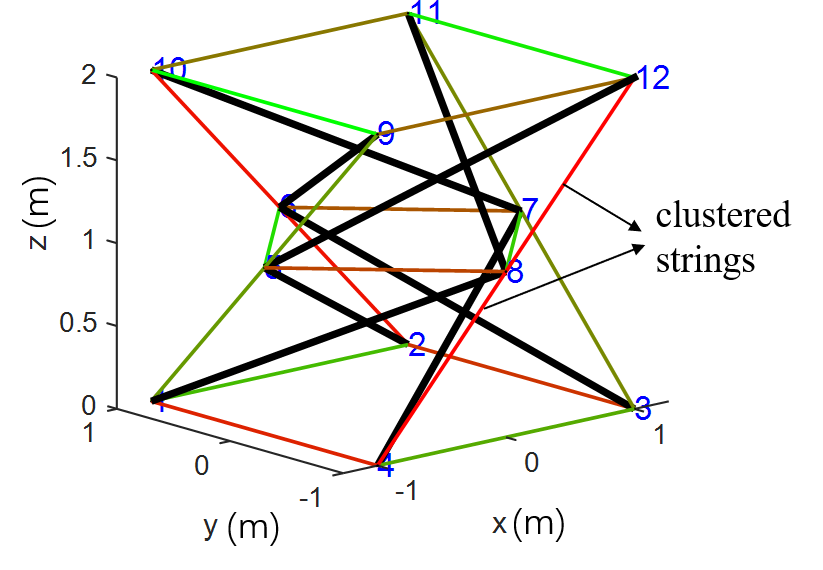}
    \caption{Configuration of a two-stage clustered tensegrity tower.}
    \label{configuration2}
\end{figure}

In this example, we analyzed a widely studied structure, a two-stage prism tower. Ali et al. \cite{ali2011analysis} and Kan et al. \cite{kan2018nonlinear} clustered the structure and investigated its quasi-static and dynamic performance of the deployment. The clustering strategy is that the vertical 8 strings into 4 ones, i.e., string from node 4 to node 8 and string from 8 to 12 are clustered into one single string. The four strings on the top side are not clustered. Note that bars are in black, strings are in other colors, and the strings being clustered are in the same color, as shown in Fig.\ref{configuration2}. And the four bottom nodes are pinned to the ground. In this paper, we implemented the same clustering method.

 
\subsubsection{Static analysis}

From the structure topology and Eq. (\ref{A_{1a}}), we know that the null space of the equilibrium matrix has five columns, which means the structure has five prestress modes. This is reasonable because four prestress modes are due to the four bottom strings pinned to the ground, and the fifth prestress mode represents the stress level of the whole above structure. We first assign the initial prestress of the four bottom strings and the vertical string to be 100N. The material properties are given in Table.\ref{the material parameters of the structure}. The cross-sectional area of bars, vertical strings, bottom strings, middle strings and top strings are $2.53\times10^{-4}{\rm m}^2, 8.17\times10^{-7}{\rm m}^2, 8.17\times10^{-7}{\rm m}^2, 1\times10^{-8}{\rm m}^2, 5.78\times10^{-7}{\rm m}^2$, respectively. Note that the cross area of all members is designed by 10\% of yielding or buckling load. The cross-area of the middle string is reduced to $1\times10^{-8}{\rm m}^2$ for the consideration of easy deployment. Because reducing the cross-sectional area of the middle strings reduces the energy required for the deployment and prevents the structure from global buckling due to high prestress levels, which is essential for a successful deployment.\par

\begin{figure}
    \centering
    \includegraphics[scale=0.6]{ 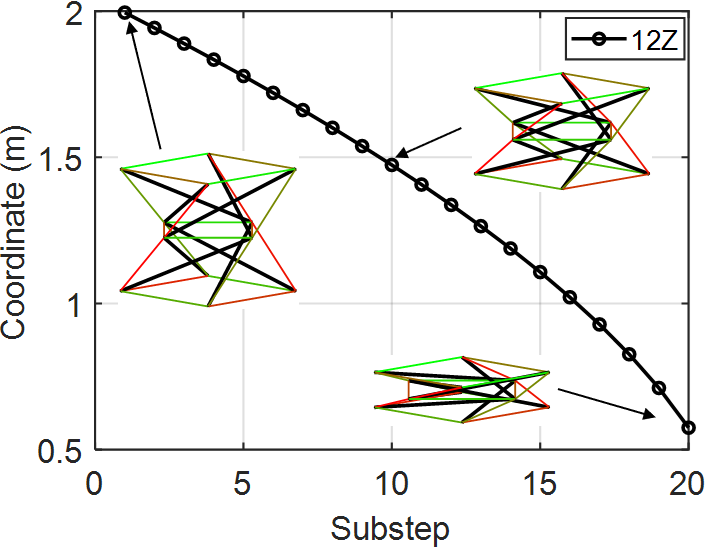}
    \caption{Structure configuration vs. substep.}
    \label{structure configuration to substep2}
\end{figure}

The structure is folded by decreasing the rest length of the four vertical strings by 0.7m simultaneously. And the length of the middle strings will increase as the height of the structure decreases. The quasi-static analysis is performed to show the feasibility of the deployment strategy. Fig.\ref{structure configuration to substep2} shows the height change of the structure in each substep. 


\subsubsection{Dynamic analysis}

\begin{figure}
    \centering
    \includegraphics[scale=0.45]{ 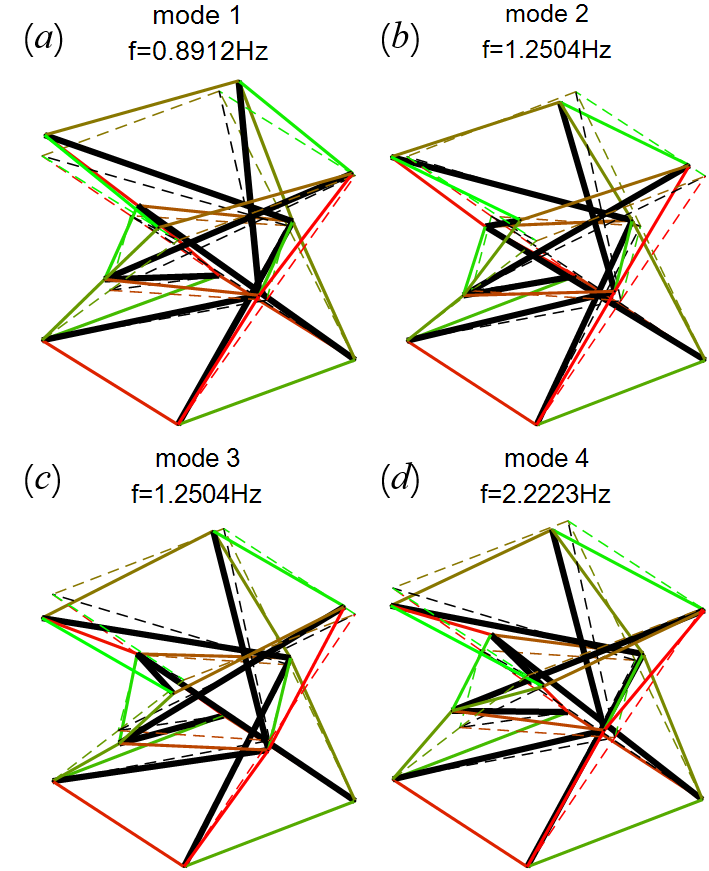}
    \caption{Free vibration mode of tensegrity tower.}
    \label{vibration mode 2}
\end{figure}

Fig.\ref{vibration mode 2} shows the natural frequencies and vibration modes. As we can see, the first mode is clockwise rotation along the Z-axis, the second and third modes are structure bending, and the fourth mode is counterclockwise rotation along the Z-axis. 
 
\begin{figure}
    \centering
    \includegraphics[scale=0.4]{ 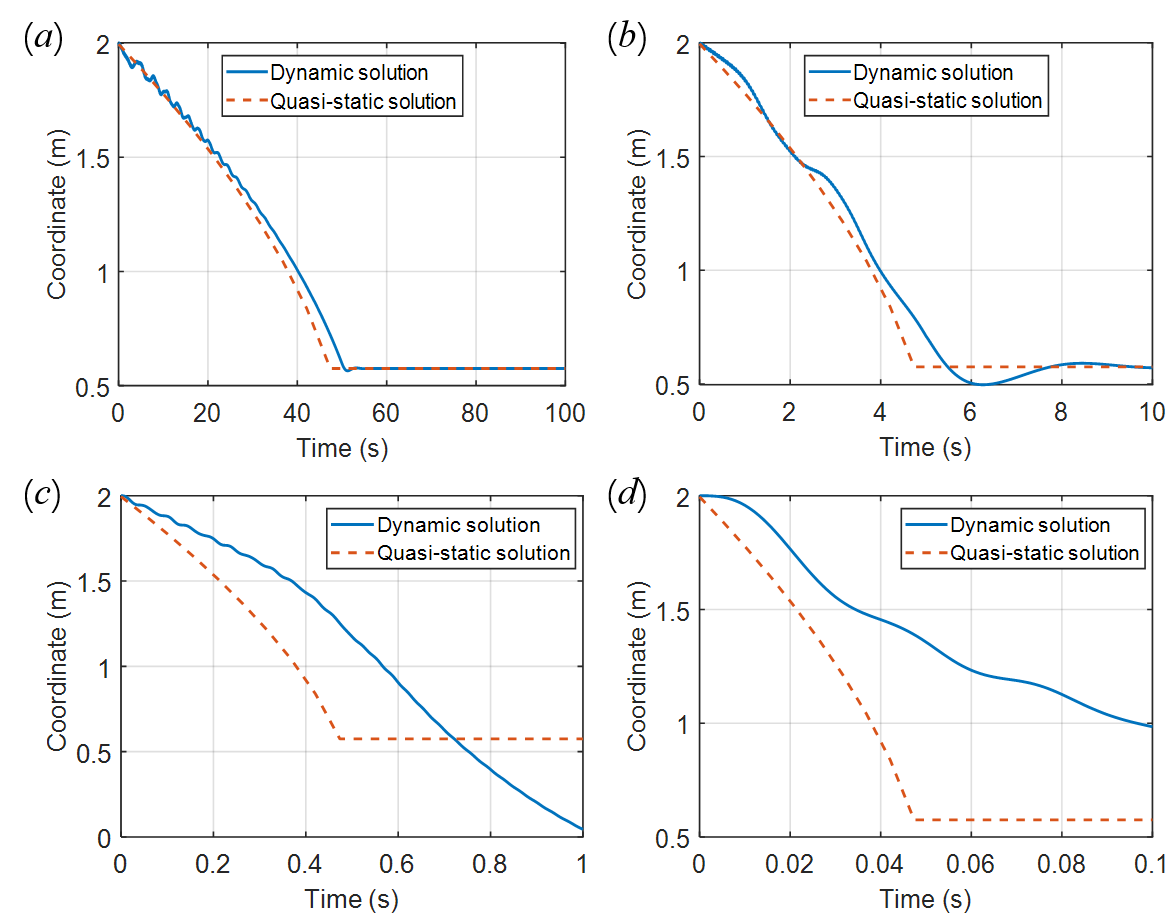}
    \caption{Time history curves of the Z-coordinate of top nodes.}
    \label{time history 2}
\end{figure}

The structure is actuated by strings in the same way as in quasi-static analysis. The dynamic analysis of the deploying process is investigated at different actuation speeds. Fig.\ref{time history 2} shows the time history of the Z-coordinate of the top node for actuation in T = 100s, 10s, 1s, 0.1${\rm s}$. We can observe that at a lower actuation speed, the dynamic response is very close to the quasi-static result. But for a fast actuation, the dynamic response shift from quasi-static dramatically. 


\begin{figure}
    \centering
    \includegraphics[scale=0.33]{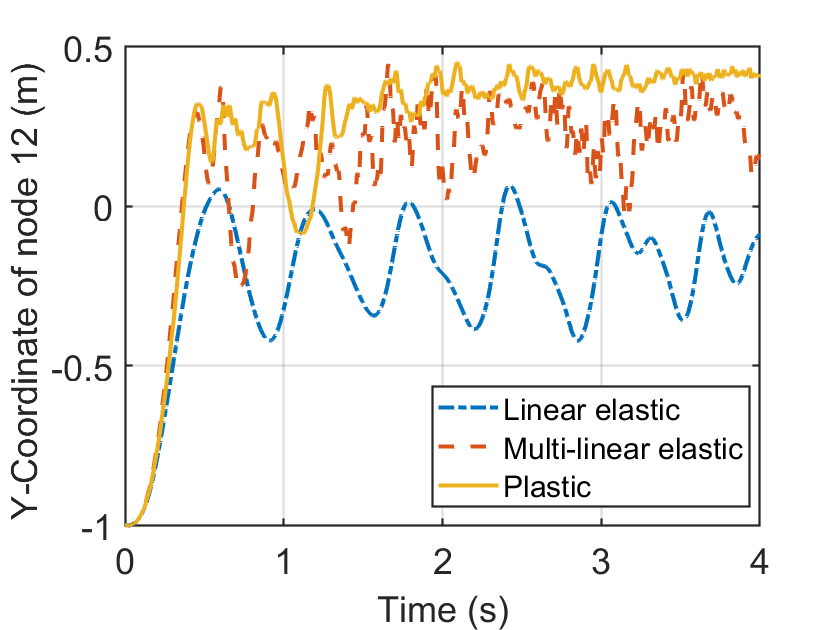}
    \caption{Time history of nodal coordinate with different material properties.}
    \label{elasto-plastic 2}
\end{figure}

Here, we also want to demonstrate the proposed dynamic equations are capable of dealing with the fem analysis with various materials. We implement a 100N force on the positive direction of the Y-axis in nodes 9 to 12. The dynamic response time histories with different materials (linear elastic, multi-linear elastic, and plastic) are studied. The Y-coordinate of node 12 is plotted in Fig. \ref{elasto-plastic 2} to show the differences.


\subsubsection{Shape control}
\begin{figure}
    \centering
    \includegraphics[scale=0.45]{ 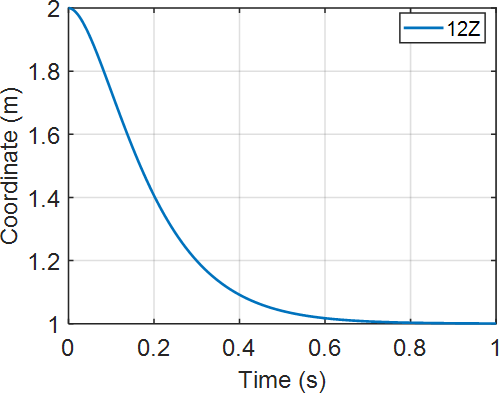}
    \caption{Time history curves of the Z-coordinate of top nodes.}
    \label{time history 2 control}
\end{figure}

\begin{figure}
    \centering
    \includegraphics[scale=0.45]{ 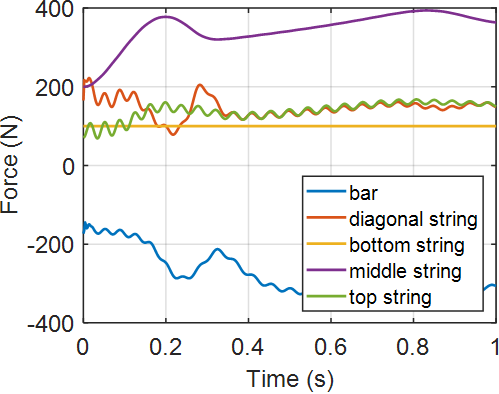}
    \caption{Time history curves of member forces.}
    \label{member force 2}
\end{figure}

The objective of shape control is to fold the tensegrity tower, where we control the Z-coordinate of nodes 9, 10, 11, and 12 from 2m to 1m. The active member is chosen as all the vertical strings. The coefficients in the control law is $\bm{\psi}=2\sqrt{100} \bm{I}$ and $\bm{\phi}=100 \bm{I}$. Fig.\ref{time history 2 control} shows the coordinate of the target nodes, and we can see that the target coordinate is driven from 2m to 1m smoothly. Fig.\ref{member force 2} shows the member force in the control process.

 \subsection{A deployable cable dome}

\begin{figure}
    \centering
    \includegraphics[scale=0.5]{ 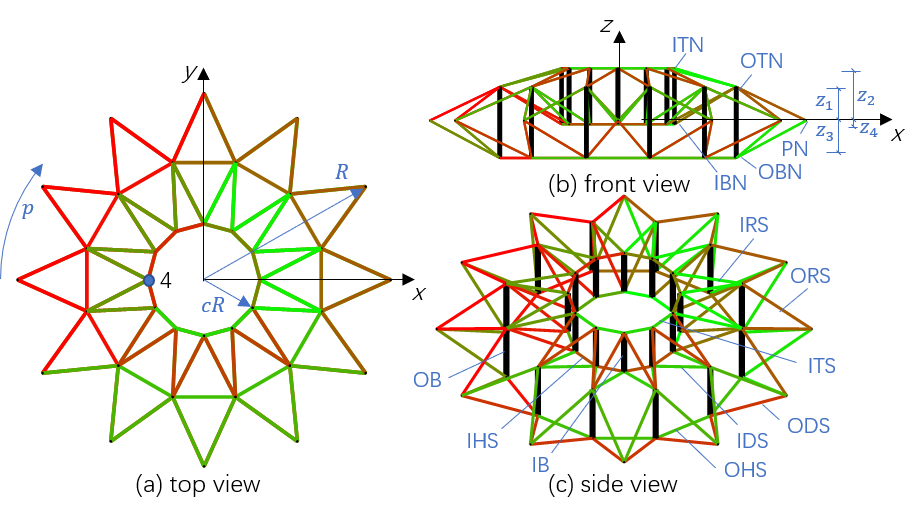}
    \caption{Configuration of a clustered Levy cable dome.}
    \label{3_configuration_Levy}
\end{figure}

This section shows a more complex example, the dynamics and control of a deployable clustered tensegrity cable dome. A Levy dome \cite{zhang2007initial}, as shown in Fig.\ref{3_configuration_Levy}, consists of nine groups of members: outer bar (OB), inner bar (IB), outer ridge strings (ORS), outer diagonal strings (ODS), inner ridge strings (IRS), inner diagonal strings (IDS), outer hoop strings (OHS), inner hoop strings (IHS), and top hoop strings (THS). Levy dome consists of five groups of nodes: outer top nodes (OTN), outer bottom nodes (OBN), inner top nodes (ITN), inner bottom nodes (IBN), and pinned nodes (PN). The deployable cable dome structure \cite{ma2021design} is modified from a Levy dome by clustering the strings in each group of members into three clustered strings. In Fig.\ref{3_configuration_Levy}, the connected strings plotted in the same color represent one single clustered string. The configuration of a Levy cable dome can be parameterized by these variables: radius of the outer ring (R), deployment ratio (c), complexity (p), Z-coordinate of the free nodes $z_1, z_2, z_3$, and $z_4$. The cable dome can be opened by increasing the rest length of all hoop strings and decreasing the rest length of the other strings.


\subsubsection{Static analysis}
\begin{figure}
    \centering
    \includegraphics[scale=0.5]{ 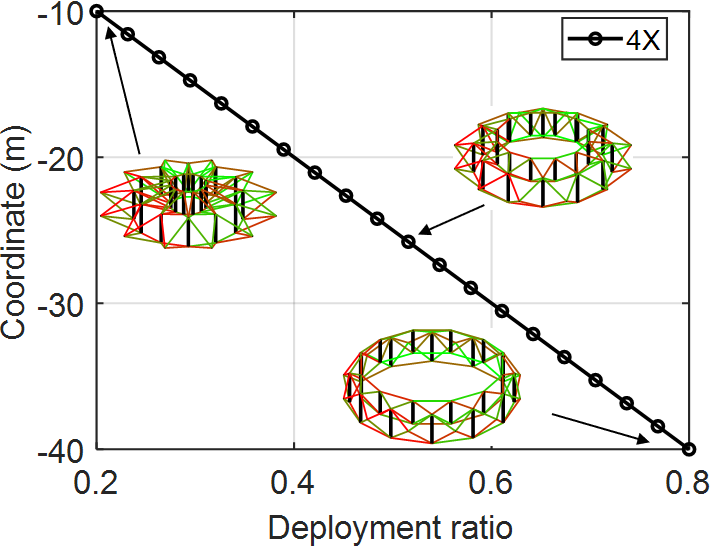}
    \caption{X-coordinate of a inner top node respect to deployment ratio.}
    \label{3_configuration to deployment ratio}
\end{figure}

\begin{figure}
    \centering
    \includegraphics[scale=0.5]{ 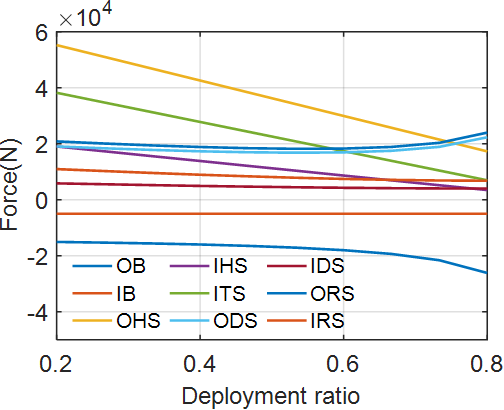}
    \caption{Member force vs. deployment ratio.}
    \label{3_static_member_force}
\end{figure}

\begin{figure}
    \centering
    \includegraphics[scale=0.5]{ 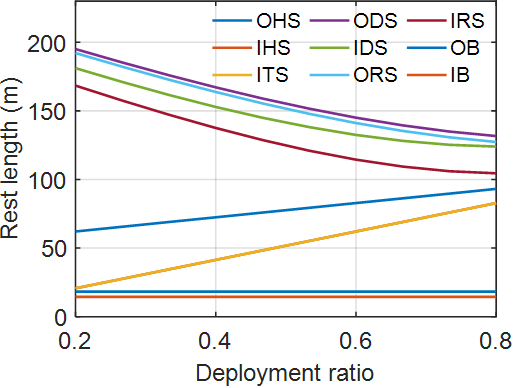}
    \caption{Rest length of clustered strings vs. deployment ratio.}
    \label{3_static_restlength}
\end{figure}

The cable dome structure has one integral feasible prestress mode, so the prestress of the structure can be determined by assigning the force of one group of the member. For example, we assign the inner vertical bars to be 5,000N in compression. The deployment ratio $c\in [0, 1]$ represents the radius of the inner hoop, and we can design the deployment trajectory by changing the parameter $c$ from 0.2 to 0.8. Fig.\ref{3_configuration to deployment ratio} shows the X-coordinate of one inner top node with respect to deployment ratios. Fig.\ref{3_static_member_force} shows the prestress of all groups of members of the cable dome in the deployment process. With the information of member force and length, we can recalculate the rest length of all the members in each configuration, as shown in Fig.\ref{3_static_restlength}.
  
\subsubsection{Dynamic analysis}

\begin{figure}
    \centering
    \includegraphics[scale=0.4]{ 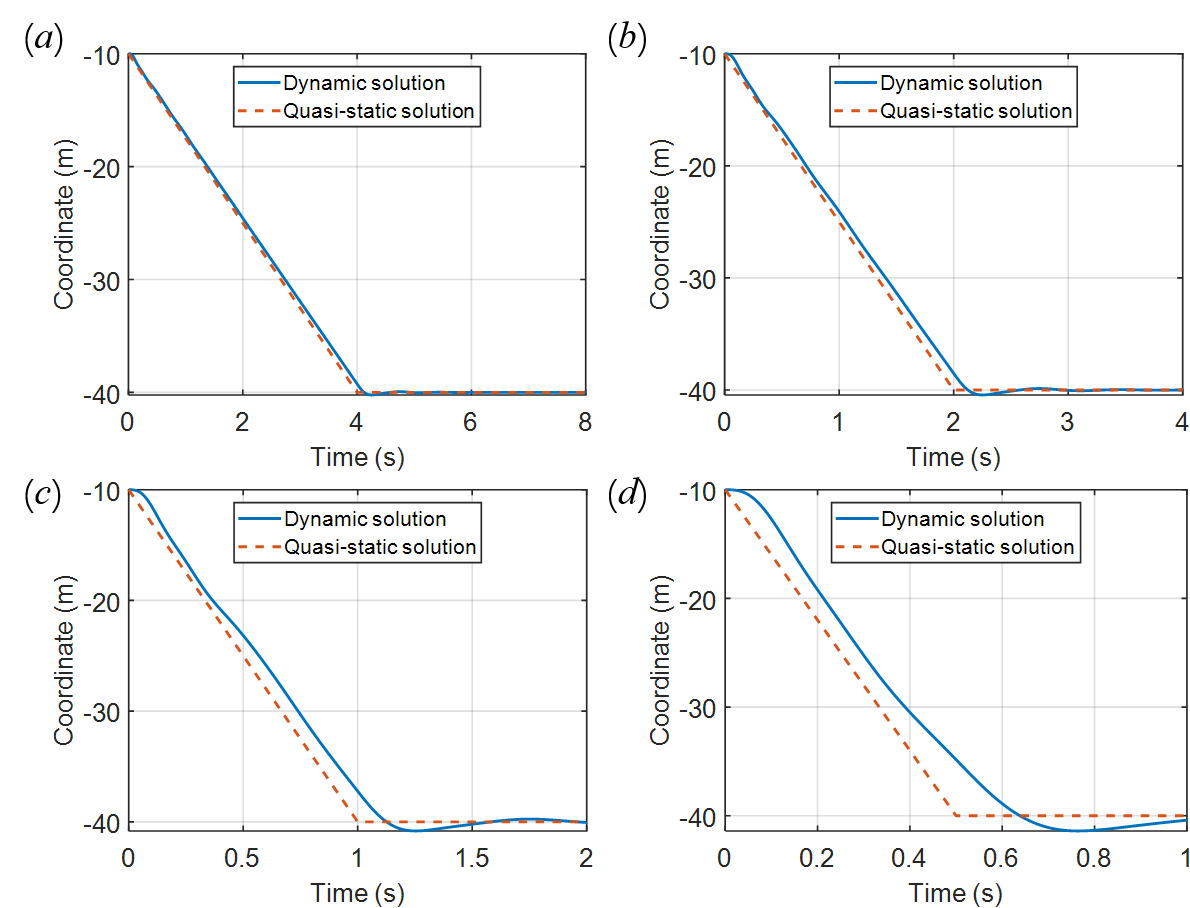}
    \caption{Time history curves of the X-coordinate of inner top node.}
    \label{3_dynamic}
\end{figure}

The structure is actuated by changing the rest length of strings, as shown in Fig.\ref{3_static_restlength}. And the dynamic analysis of the deployment process at different speeds is investigated. Fig.\ref{3_dynamic} shows the time history of the X-coordinate of the inner top node for a total deployment time $T=8$s, $4$s, $2$s, and $1{\rm s}$. In this period, in the first half part, we actuate the strings, while in the second half, we stop the actuation and leave the system free response. One can observe that the dynamic response with lower actuation speeds is closer to the quasi-static result. 


\subsubsection{Shape control}

\begin{figure}
    \centering
    \includegraphics[scale=0.4]{ 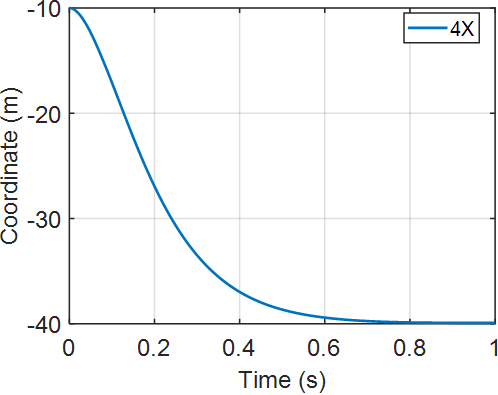}
    \caption{Time history curves of the X-coordinate of a inner top node.}
    \label{3_control_coordinate}
\end{figure}

\begin{figure}
    \centering
    \includegraphics[scale=0.45]{ 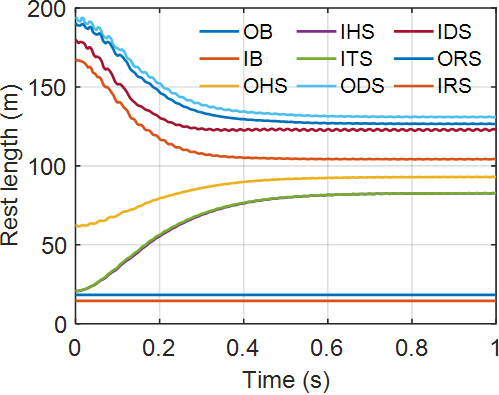}
    \caption{Control of rest length with respect to time.}
    \label{3_control_rest_length}
\end{figure}

The objective of shape control is to let the cable dome deploy from the configuration with a deployment ratio of $\textit{c}=0.2$ to $\textit{c}=0.8$. All the nodal coordinates of free nodes are chosen as the control targets. All the clustered strings are set as active members, and all the bars are passive members. 
The coefficients in the error dynamic are  $\bm{\psi}=2\sqrt{100}\bm{I}$ and $\bm{\phi}=100\bm{I}$. We can see that the tendency of the rest length in control is similar to that in the quasi-static analysis, as shown in  Fig.\ref{3_static_restlength}. The movement of the target node is much more smooth compared with the results of the dynamics, as shown in Fig.\ref{3_dynamic}. Fig.\ref{3_control_coordinate} shows the time history of the X-coordinate of node 4 (an inner top node). The control strategy of the rest length of strings is shown in Fig.\ref{3_control_rest_length}. 


\section{Conclusion}\label{section8}


The clustered tensegrity has shown many advantages in structure morphing, i.e., require fewer actuators, fewer sensors, and less control efforts. There have been only limited investigations from a quasi-static or dynamic point of view in recent years. This paper derives a systematic analytical dynamic equation of clustered tensegrity and a shape control law. The nonlinear clustered tensegrity dynamics is developed based on the finite element analysis approach and Lagrangian method with nodal coordinate vector as the variable. The mass, damping, stiffness, and tangent stiffness matrices are explicitly obtained and explained. This approach allows one to conduct comprehensive studies on any CTS with any node constraints and various load conditions. The dynamic deployment analysis shows that the dynamic solution differs from the quasi-static process as the actuation speed increases.  And the control examples show that shape control law is able to move the tensegrity structure to the desired configuration smoothly. \par

\section{Acknowledgments}
The first author is grateful for the financial support by the Foundation of Key Laboratory of Space Structures of Zhejiang Province (No.202102).

\printcredits

\bibliographystyle{cas-model2-names}

\bibliography{cas-refs}
\end{sloppypar}

\end{document}